\documentstyle[12pt, hyperref]{article}

\catcode`\@=11
\def\theequation{\thesubsection.\arabic{equation}}
\@addtoreset{footnote}{section}
\@addtoreset{footnote}{subsection}
\@addtoreset{equation}{subsection}
\catcode`\@=12
\catcode`\@=11

\newcounter{app}

\def\app{\setcounter{equation}{0}
\def\theequation{A\arabic{app}.\arabic{equation}}\par
   \addvspace{4ex}
   \@afterindentfalse
  \secdef\@app\@dapp}
\newcommand\@app{\@startsection {app}{1}{0ex}%
                                   {-3.5ex \@plus -1ex \@minus -.2ex}%
                                   {2.3ex \@plus.2ex}%
                                   {\normalfont\Large\bf}}

\def\@dapp#1{%
{\parindent \z@ \raggedright  \bf #1}\par\nobreak}
\def\l@app#1#2{\ifnum \c@tocdepth >\z@
    \addpenalty\@secpenalty
    \addvspace{1.0em \@plus\p@}%
    \setlength\@tempdima{8.5em}%
    \begingroup
      \parindent \z@ \rightskip \@pnumwidth
      \parfillskip -\@pnumwidth
      \leavevmode \bfseries
      \advance\leftskip\@tempdima
      \hskip -\leftskip
      #1\nobreak\hfil \nobreak\hb@xt@\@pnumwidth{\hss #2}\par
    \endgroup\fi}
\catcode`\@=12

\small  \oddsidemargin -10mm  \evensidemargin 5mm  \topmargin -10mm
\def\be{\begin{equation}}
\def\ee{\end{equation}}
\def\g{\gamma}

\renewcommand{\l}{\langle}
\renewcommand{\r}{\rangle}

\newtheorem{prop}{Proposition}
\newtheorem{remark}{Remark}

\def\bprop{\begin{prop}}
\def\eprop{\end{prop}}
\def\bremark{\begin{remark}}
\def\eremark{\end{remark}}
\begin{document}
\begin{center}
{\large\bf Fermionic representation for basic hypergeometric functions\\
 related to Schur polynomials}\\
\vspace{.5cm}
%\vspace{1cm}
{\bf A. Yu. Orlov\footnote{E-mail: orlovs@kusm.kyoto-u.ac.jp and orlovs@wave.sio.rssi.ru}}
\vspace{.2cm}\\
   Department of Mathematics, Faculty of Science, \\Kyoto University,
Kyoto 606-85-02, Japan
\vspace{.1cm}\\
permanent address: Nonlinear wave processes laboratory,\\
Oceanology Institute, 36 Nakhimovskii prospekt\\
Moscow 117851, Russia.
\vspace{.5cm}\\
{\bf  D. M. Scherbin\footnote{E-mail:
 scherbin@wave.sio.rssi.ru}},
\vspace{.2cm}\\
Nonlinear wave processes laboratory,\\
Oceanology Institute, 36 Nakhimovskii prospekt\\
Moscow 117851, Russia.\\
\end{center}
%\vspace{1.5cm}
\begin{abstract}
We present the fermionic representation for the $q$-deformed hypergeometric
functions related to Schur polynomials considered by S.Milne \cite{Milne}. 
For $q=1$ these functions are also known as hypergeometric functions
of matrix argument which are related to zonal spherical polynomials
for $GL(N,C)/U(N)$ symmetric space. We show that these multivariable 
hypergeometric functions are tau-functions of the KP hierarchy.
At the same time they are the ratios of Toda lattice tau-functions 
considered by Takasaki in \cite{Tinit}, \cite{T} evaluated at certain
values of higher Toda lattice times.  The variables of the 
hypergeometric functions are related to the higher times of those  
hierarchies via Miwa change of variables. The discrete Toda lattice variable 
shifts parameters of hypergeometric functions.
Hypergeometric functions of type ${}_pF_s$ can be also viewed as
group 2-cocycle for the $\Psi$DO on the circle of the order $p-s \leq 1$ 
(the group times are higher times of TL hierarchy and the arguments of
hypergeometric function). 
We get the determinant representation and the integral representation of 
special type of KP tau-functions, these results generalize some of Milne's
results in \cite{Milne}. We
write down a system of linear differential and difference equations for these
tau-functions (string equations). We present also fermionic representation
for special type of Gelfand-Graev hypergeometric functions.
\end{abstract}
\tableofcontents
\section*{Introduction}
Hypergeometric functions play an important role both, in physics and in
mathematics \cite{GGR}. Many special functions and polynomials 
(such as $q$-Askey-Wilson polynomials, $q$-Jacobi polynomials,
$q$-Gegenbauer polynomials , $q$-Racah polynomials , 
$q$-Hahn polynomials , expressions for Clebsch-Gordan coefficients) 
are just
certain hypergeometric functions evaluated at special values of parameters.
In physics hypergeometric functions and their $q$-deformed counterparts 
sometimes play the role of wave functions and
correlation functions for quantum integrable systems.
In the present paper we shall construct hypergeometric functions as
 tau-functions $\tau$ of the Kadomtsev-Petviashvili (KP) hierarchy of 
equations. It is interesting that the KP equation
\be\label{KP}
4\partial_{t_1}\partial_{t_3}u=\partial_{t_1}^4 u+
3\partial_{t_2}^2u+3\partial_{t_1}^2 u^2 \quad 
(u=2\partial_{t_1}^2\log \tau),
\ee
which originally served in plasma physics \cite{KP} now plays a very 
important role both, in physics (see \cite{ZM}; see review in \cite{M}
 for modern applications) and in mathematics. 
The peculiarity of Kadomtsev-Petviashvili
equation appeared in the paper \cite{Dr} where L-A pair of KP
equation was presented, and mainly in the paper of V.E.Zakharov and 
A.B.Shabat in 1974
where this equation was integrated by the dressing method. Actually it
was the paper \cite{ZSh} where so-called hierarchy of higher KP 
equations appeared.
 Another very important equation is the two-dimensional Toda lattice
(TL) integrated first in \cite{AM}. 
In the present paper we use these equations to construct hypergeometric
functions which depend on many variables, these variables are KP and
Toda lattice higher times. 
Here we shall use the general approach to integrable hierarchies of Kyoto 
school \cite{DJKM}, see
 also \cite{JM,D}. Especially a set
of papers about Toda lattice \cite{UT,Tinit,TI,TII,TT,NTT,T} is important for
us.

About the structure of the paper. To learn the main result, which is the 
fermionic representation of the hypergeometric functions, one needs
read only {Sections 1} and Subsections 2.1 and 2.2 - to learn the 
notations, and then read Subsection 3.1, and {Examples 3-6}. The 
linear equations (constraints) 
for the tau-function, which generalize familiar Gauss equation for 
the well-known Gauss hypergeometric function see in Subsection 3.7. 
For determinant 
representation and integral representation see Subsections 3.8 and 3.9. 
For hypergeometric function as group two-cocycle see the {Remark} in 
the  end of Appendix {``Gauss factorization problem etc.''}. All 
other material is just setting the topic into the theory of integrable 
systems.\\

Few words about notations. 
The symbols $*$ and  ${\bar{~}}$  do not denote the 
complex conjugation.
Symbol $'$ does not denote the derivative.
Bold ${\bf n}$ stands for partitions.
Bold ${\bf t}$ and ${\bf t}^*$ stand for collections of KP and TL 
higher time variables.
Bold ${\bf x}_{(N)}$ and ${\bf y}_{(N)}$ stand for Miwa variables.

\section{Milne's hypergeometric series}
\subsection{Ordinary hypergeometric functions}
First let us remember that generalized hypergeometric function of
one variable $x$ is defined as
\be\label{ordinary}
{}_pF_s\left(a_1,\dots ,a_p; b_1,\dots ,b_s; x \right)
=\sum_{n=0}^\infty\frac {(a_1)_n \cdots (a_p)_n}
             {(b_1)_n \cdots (b_s)_n}
\frac{x^n}{n!} .
\ee
Here $(a)_n$ is Pochhammer's symbol:
\be
(a)_n=\frac {\Gamma(a+n)}{\Gamma(a)}=a(a+1)\cdots (a+n-1) .
\ee
Given number $q$, $|q|<1$, the so-called basic hypergeometric series of one
variable is defined as
\be\label{qordinary}
{}_p\Phi_s\left(a_1,\dots ,a_p; b_1,\dots ,b_s; q, x \right)
=\sum_{n=0}^\infty \frac {(q^{a_1};q)_n \cdots (q^{a_p};q)_n}
             {(q^{b_1};q)_n \cdots (q^{b_s};q)_n}
\frac{x^n}{(q;q)_n} .
\ee
Here $(q^a,q)_n$ is $q$-deformed  Pochhammer's symbol:
\be\label{qPochh}
(q^a;q)_0=1,\quad (q^a;q)_n=(1-q^a)(1-q^{a+1})\cdots(1-q^{a+n-1}) .
\ee
Both series converge for all $x$ in case $p<s+1$. 
In case $p=s+1$ they converge for $|x|<1$. We refer these 
well-known hypergeometric
functions as ordinary hypergeometric functions.
\subsection{The multiple
basic hypergeometric series related to Schur polynomials}
There are several well-known different multivariable generalizations of 
hypergeometric series of one variable \cite{V, KV}. 
Let $|q|<1$ and let ${\bf x}_{(N)}=(x_1,\dots,x_N)$ be indeterminates. Let 
$s_{\bf n}(x_1,x_2,...,x_N)$ be the Schur polynomial corresponding to
a partition ${\bf n}$ \cite{Mac}. $s_{\bf n}(x_1,x_2,...,x_N)$ is
a symmetric function of variables $x_k$.
The multiple
basic hypergeometric series related to Schur polynomials were introduced 
by S.Milne \cite{Milne} as
\begin{eqnarray}\label{vh}
{}_p\Phi_s\left(a_1,\dots ,a_p;b_1,\dots ,b_s;q,{\bf x}_{(N)}\right)
=\sum_{{\bf n}\atop l({\bf n})\le N}\frac {(q^{a_1};q)_{{\bf n}}
\cdots (q^{a_p};q)_{{\bf n}}}
{(q^{b_1};q)_{{\bf n}}\cdots (q^{b_s};q)_{{\bf n}}}
\frac{q^{n({\bf n})}}{H_{{\bf n}}(q)}
s_{{\bf n}} \left( {\bf x}_{(N)} \right) ,
\end{eqnarray}
where the sum is over all different partitions 
${\bf n}=\left(n_1,n_2,\dots,n_r\right)$, where 
$n_1\ge n_2\ge \cdots \ge n_r$, $r\le |{\bf n}|$, 
$|{\bf n}|=n_1+\cdots+n_r$ and whose length $l({\bf n})=r\le N$. 
Schur polynomial  $s_{{\bf n}} \left( {\bf x}_{(N)} \right)$, with 
$N \ge l({\bf n})$, 
is a symmetric function of variables
${\bf x}_{(N)}$ and  defined as follows \cite{Mac}:
\be\label{Schur}
s_{{\bf n}}({\bf x}_{(N)})=\frac{a_{{\bf n}+\delta}}{a_{\delta}},\quad
a_{{\bf n}}=\det (x_i^{n_j})_{1\le i,j\le N},\quad \delta=
(N-1, N-2,\dots,1,0).
\ee
Coefficient $(q^c;q)_{\bf n}$ associated with partition ${\bf n}$
 is expressed in terms of the $q$-deformed Pochhammer's Symbols 
$(q^c;q)_{n}$ (\ref{qPochh}):
\be
(q^c;q)_{\bf n}=(q^c ;q)_{n_1}(q^{c-1};q)_{n_2}\cdots (q^{c-l+1};q)_{n_l}.
\ee
The multiple $q^{n({\bf n})}$ defined on the partition ${\bf n}$: 
\begin{equation}\label{n(n)}
q^{n({\bf n})}=q^{\sum_{i=1}^N (i-1)n_i} ,
\end{equation}
and $q$-deformed 'hook polynomial' $H_{{\bf n}}(q)$ is 
\begin{eqnarray}\label{hp}
H_{{\bf n}}(q)=\prod_{(i,j)\in {\bf n}} \left(1-q^{h_{ij}}\right) , \quad
h_{ij}=(n_i+n'_j-i-j+1) ,
\end{eqnarray}
where ${\bf n}'$ is the conjugated
partition (for the definition see \cite{Mac}).
For $N=1$ we get (\ref{qordinary}). \\
Another generalization of hypergeometric series is so-called 
hypergeometric function of matrix argument ${\bf X}$ with 
indices $\bf{a}$ and $\bf{b}$ \cite{V}: 
\begin{eqnarray}\label{hZ1}
{}_pF_s\left.\left(a_1,\dots ,a_p\atop b_1,\dots ,b_s\right|{{\bf{X}}}
\right)
=\sum_{{\bf n}}\frac {(a_1)_{{\bf n}} \cdots (a_p)_{{\bf n}}}
                     {(b_1)_{{\bf n}} \cdots (b_s)_{{\bf n}}}
\frac{Z_{\bf n}({\bf X})}
{|{\bf n}|!} .
\end{eqnarray}
Here ${\bf X}$ is a Hermitian $N\times N$ matrix, and $Z_{\bf n}({\bf X})$
is zonal spherical polynomial for the symmetric space $GL(N,C)/U(N)$, 
see \cite{V, KV}. 
Let us note that
in the limit $q \to 1$ series (\ref{vh}) 
coincides with (\ref{hZ1}), see \cite{KV}.\\

\subsection{Hypergeometric series of double set of arguments}
The formula
\[
{}_p\Phi_s\left.\left(a_1,\dots ,a_p\atop b_1,\dots ,b_s\right|q,
{\bf x}_{(N)},{\bf y}_{(N)}\right)=
\]
\be
\sum_{{\bf n}\atop l({\bf n})\le N}\frac {(q^{a_1};q)_{{\bf n}}\cdots (q^{a_p};q)_{{\bf n}}}
                            {(q^{b_1};q)_{{\bf n}}\cdots (q^{b_s};q)_{{\bf n}}}
\frac{q^{n({\bf n})}}{H_{{\bf n}}(q)}
\frac
{s_{{\bf n}} \left( {\bf x}_{(N)} \right)s_{{\bf n}} \left( {\bf y}_{(N)} \right)}
{s_{{\bf n}} \left(1,q,q^2,...,q^{N-1} \right)} \label{vh2}
\ee
defines the multiple basic hypergeometric function of two sets of variables
which was also introduced by S.Milne, see \cite{Milne}, \cite{KV}.
\\
Another generalization of hypergeometric series is so-called 
hypergeometric function of matrix arguments ${\bf X},{\bf Y}$ with 
indices $\bf{a}$ and $\bf{b}$: 
\begin{eqnarray}\label{hZ}
{_p{\mathcal{F}}_s}\left(a_1,\dots ,a_p;b_1,\dots ,b_s;{{\bf{X}}}, 
{{\bf{Y}}}\right)
=\sum_{{\bf n}}\frac {(a_1)_{{\bf n}} \cdots (a_p)_{{\bf n}}}
                     {(b_1)_{{\bf n}} \cdots (b_s)_{{\bf n}}}
\frac{Z_{{\bf n}}({\bf X})Z_{{\bf n}}({\bf Y})}
{|{\bf n}|!Z_{{\bf n}}({\bf I}_n)} .
\end{eqnarray}
Here ${\bf X},{\bf Y}$ are Hermitian $N\times N$ matrices and $Z_{{\bf n}}({\bf X}), Z_{{\bf n}}({\bf Y})$
are zonal spherical polynomials for the symmetric spaces
$GL(N,C)/U(N)$, $GL(N,R)/SO(N)$ and $GL(N,H)/Sp(N)$ 
 see \cite{KV}. In our paper we shall consider only the first case;
different hypergeometric functions related to zonal polynomials for 
symmetric spaces $GL(N,C)/U(N)$, $GL(N,R)/SO(N)$ and 
$GL(N,H)/Sp(N)$ \cite{KV} will not be considered.

There are also hypergeometric functions related to Jack polynomials 
$C_{\bf n}^{(d)}$ \cite{KV}:
\begin{eqnarray}\label{hJ}
{_p{\mathcal{F}}_s}^{(d)}\left(a_1,\dots ,a_p;b_1, \dots b_s;{\bf x}_{(N)},
{\bf y}_{(N)} \right)=\nonumber\\
\sum_{{\bf n}}\frac {(a_1)^{(d)}_{{\bf n}}\cdots (a_p)^{(d)}_{{\bf n}}}
                    {(b_1)^{(d)}_{{\bf n}}\cdots (b_s)^{(d)}_{{\bf n}}}
\frac{C_{\bf n}^{(d)}({\bf x}_{(N)})
C_{\bf n}^{(d)}({\bf y}_{(N)})}
{|{\bf n}|!C_{\bf n}^{(d)}(1^n)} ,
\end{eqnarray}
where
\begin{equation}\label{ad}
(a)^{(d)}_{{\bf n}}=\prod_{i=1}^{l({\bf n})}
\left(a-\frac{d}{2}(i-1)\right)_{n_i} .
\end{equation}
Here $(c)_k=c(c+1)\cdots (c+k-1)$. It is known that
for the special value $d=2$ the last expression (\ref{hJ}) coincides 
with (\ref{vh}), and coincides with (\ref{hZ}) as $|q| \to 1$. These 
last cases we shall consider below.

\section{A brief introduction to the fermionic description of the 
KP and TL hierarchies \cite{DJKM, JM, UT}}
\subsection{Fermionic operators and Fock space}
We have fermionic fields:
\be \label{fermions}
\psi(z)=\sum_k \psi_k z^k ,\qquad \psi^*(z)=\sum_k \psi^*_k z^{-k-1}dz ,
\ee
where fermionic operators satisfy the canonical anti-commutation relations:
\be \label{antikom}
[\psi_m,\psi_n]_+=[\psi^*_m,\psi^*_n]_+=0;\qquad [\psi_m,\psi^*_n]_+=
\delta_{mn} .
\ee
Let us introduce left and right vacuums by the properties:
\begin{eqnarray}\label{vak}
\psi_m |0\r=0 \qquad (m<0),\qquad \psi_m^*|0\r =0 \qquad (m \ge 0) , \\
\l 0|\psi_m=0 \qquad (m\ge 0),\qquad \l 0|\psi_m^*=0 \qquad (m<0) .
\end{eqnarray}
The vacuum expectation value is defined by relations:
\begin{eqnarray}\label{psipsi*vac} 
\l 0|1|0 \r=1,\quad \l 0|\psi_m\psi_m^* |0\r=1\quad m<0, 
\quad  \l 0|\psi_m^*\psi_m |0\r=1\quad m\ge 0 ,
\end{eqnarray}
\be\label{end}
 \l 0|\psi_m\psi_n |0\r=\l 0|\psi^*_m\psi^*_n |0\r=0,\quad \l 
0|\psi_m\psi_n^*|0\r=0 \quad m\ne n .
\ee
Let us notice that relations (\ref{antikom})-(\ref{end}) 
are invariant under the  transformation
\begin{equation}\label{gauge}
\psi_n \to e^{-T_n}\psi_n, 
\qquad \psi^*_n \to e^{T_n}\psi^*_n \qquad (T_n \in C) .
\end{equation}

Consider infinite matrices $(a_{ij})_{i,j\in Z}$ satisfying the condition:
there exists an $N$ such that  $a_{ij}=0$ for $|i-j|>N$.
Let us take the set of linear combinations of quadratic elements 
$\sum a_{ij}:\psi_i\psi_j^*:$, where $: :$ means the normal ordering 
$:\psi_i\psi_j^*:=\psi_i\psi_j^*-\l 0|\psi_i\psi_j^*|0\r$.  These elements 
together with $1$ span an infinite dimensional Lie algebra 
$\widehat{gl}(\infty)$:
\be\label{commutator} 
[\sum a_{ij}:\psi_i \psi_j^*:,\sum b_{ij} :\psi_i\psi_j^*:]=
\sum c_{ij}:\psi_i \psi_j^*:+c_0 ,
\ee
\be
c_{ij}=\sum_k a_{ik}b_{kj}-\sum_k b_{ik}a_{kj} ,
\ee
\be\label{cocycle}
c_0=\sum_{i<0,j\ge 0} a_{ij}b_{ji}-\sum_{i\ge 0,j<0} a_{ij}b_{ji} .
\ee
Now we define the operator $g$ which is an element of the group corresponding 
to the Lie algebra $\widehat{gl}(\infty)$:
\be
g\psi_ng^{-1}=\sum_m \psi_m a_{mn},\qquad g^{-1}\psi^*_n g=
\sum_m a_{nm}\psi^*_m .
\label{spin}
\ee
\subsection{The KP and Toda tau functions}
First let us define the vacuum vectors labeled by the integer $M$:
\begin{eqnarray}
\l M|=\l 0|\Psi^{*}_{M},\qquad |M\r=\Psi_{M}|0\r ,
\end{eqnarray}
\begin{eqnarray}
\Psi_{M}=\psi_{M-1}\cdots\psi_1\psi_0 \quad M>0,\qquad \Psi_{M}=\psi^{*}_{M}\cdots\psi^{*}_{-2}\psi^{*}_{-1}\quad M<0 , \nonumber\\
\Psi^{*}_{M}=\psi^{*}_{0}\psi^{*}_1\cdots\psi^{*}_{M-1} \quad M>0,\qquad \Psi^{*}_{M}=\psi_{-1}\psi_{-2}\cdots\psi_{M}\quad M<0 .
\end{eqnarray}
The tau-function of the KP equation and the tau-function 
of the two-dimensional Toda lattice (TL) sometimes are defined as
\be\label{taucorKP}
\tau_{KP}(M,{\bf t})=
\langle M|e^{H({\bf t})}g|M \rangle ,
\ee
\be\label{taucor}
\tau_{TL}(M,{\bf t},{\bf t}^*)=
\langle M|e^{H({\bf t})}ge^{H^*({\bf t}^*)}|M \rangle .
\ee
According to \cite{JM} the integer $M$ in (\ref{taucor}) plays the role 
of discrete Toda lattice variable.

The times ${\bf t}=(t_1,t_2,\dots)$ and  ${\bf t}^*=(t_1^*,t_2^*,\dots)$ 
 are called higher Toda lattice times \cite{JM,UT} (the 
first set ${\bf t}$ is in the same time the set of higher KP times. 
The first times of this set $t_1,t_2,t_3$ are independent variables for 
KP equation (\ref{KP}), which is the first nontrivial equation in the 
KP hierarchy). $H({\bf t})$ and $H^*({\bf t}^*)$ 
belong to the following ${\widehat {gl}}(\infty)$ Cartan subalgebras:
\be\label{hamiltonians}
H({\bf t})=\sum_{n=1}^{+\infty} t_n H_n ,\quad 
H^*({\bf t}^*)=\sum_{n=1}^{+\infty} t_n^* H_{-n},\quad 
H_n=\frac{1}{2\pi i}\oint :z^{n}\psi(z)\psi^*(z): .
\ee
For the Hamiltonians we have Heisenberg algebra commutation relations:
\be
[H_n,H_m]=n\delta_{m+n,0} .
\ee
The action of  $e^{H({\bf t})}$ on the fermions:
\begin{eqnarray}
e^{H({\bf t})}\psi_ie^{-H({\bf t})}=
\sum_{n=0}^{+\infty} p_n({\bf t}) \psi_{i-n},\quad
e^{H({\bf t})}\psi^*_i e^{-H({\bf t})}=
\sum_{n=0}^{+\infty} p_n({\bf t}) \psi^*_{i+n} ,
\end{eqnarray}
where $p_n$ is the elementary Schur polynomial defined by the Taylor's 
expansion:
\be 
e^{\xi({\bf t},z)}=\exp(\sum_{k=1}^{+\infty}t_kz^k)=
\sum_{n=0}^{+\infty}z^n p_n({\bf t}) .
\ee  
The action on fermionic fields is especially simple:
\be
 e^{H({\bf t})}\psi(z) e^{-H({\bf t})}= \psi(z) e^{\xi({\bf t},z)},
\quad e^{H({\bf t})}\psi^*(z) e^{-H({\bf t})}
= \psi^*(z) e^{-\xi({\bf t},z)} ,
\ee
\be
 e^{-H^*({\bf t}^*)}\psi(z) e^{H^*({\bf t}^*)}= 
\psi(z) e^{-\xi({\bf t}^*,z^{-1})},
\quad e^{-H^*({\bf t}^*)}\psi^*(z) e^{H^*({\bf t}^*)}= 
\psi^*(z) e^{\xi({\bf t}^*,z^{-1})} .
\ee

In the KP theory it is suitable to use another definition of  
 the  Schur function corresponding to the partition 
${\bf n}=(n_1,\dots ,n_r)$:
\be
s_{{\bf n}}({\bf t})=\det(p_{n_i-i+j}({\bf t}))_{1\le i,j\le r} ,
\ee
where $p_m({\bf t})$ is the elementary Schur polynomial defined by the Taylor's expansion:
\be\label{elSchur} 
e^{\xi({\bf t},z)}=\exp(\sum_{k=1}^{+\infty}t_kz^k)=
\sum_{n=0}^{+\infty}z^n p_n({\bf t}) .
\ee  
It is related to $s_{{\bf n}} \left( {\bf x}_{(N)} \right)$ and $s_{{\bf n'}} \left( {\bf x}_{(N)} \right)$,
where  a partition ${\bf n'}$ is conjugated to ${\bf n}$, as follows 
\cite{Mac}:
\be
s_{{\bf n}}({\bf t^{+}}({\bf x}_{(N)}))=s_{{\bf n}} \left( {\bf x}_{(N)} \right),\quad 
s_{{\bf n}}({\bf t^{-}}({\bf x}_{(N)}))=s_{{\bf n'}} \left( {\bf x}_{(N)} \right)
\ee
via the changes of variables (which is known as Miwa change of
variables in the literature on the integrable systems):
\be\label{Miwa+}
t^{+}_m({\bf x}_{(N)})=\sum_{i=1}^{N} \frac{x_i^m}{m},
\ee
\be\label{Miwa-}
\qquad t^{-}_m({\bf x}_{(N)})=-\sum_{i=1}^{N} \frac{x_i^m}{m} .
\ee
Let us notice that $s_{{\bf n}}({\bf t^{+}}({\bf x}_{(N)}))=0$
for $l({\bf n})>N$, and $s_{{\bf n}}({\bf t^{-}}({\bf x}_{(N)}))=0$
for $l({\bf n}')>N$.

{\bf Lemma 1} \cite{DJKM} \\
For $-j_1<\cdots <-j_k<0\le i_s<\cdots <i_1$, $s-k\ge 0$ the next formula is valid:
\be\label{glemma}
\l s-k|e^{H({\bf t})}\psi^*_{-j_1}\cdots
\psi^*_{-j_k}\psi_{i_s}\cdots\psi_{i_1}|0\r=
(-1)^{j_1+\cdots +j_k+(k-s)(k-s+1)/2}s_{{\bf n}}({\bf t}) ,
\ee
where the partition ${\bf n}=(n_1,\dots , n_{s-k}, n_{s-k+1},\dots , n_{s-k+j_1})$ is defined by the pair of partitions:
\begin{eqnarray}
(n_1, \dots , n_{s-k})=(i_1-(s-k)+1, i_2-(s-k)+2, \dots , i_{s-k}) ,\\
(n_{s-k+1},\dots , n_{s-k+j_1})=(i_{s-k+1},\dots , i_s|j_1-1,\dots , j_k-1) .
\end{eqnarray}
The proof is achieved by direct calculation.
Here $(\dots | \dots)$ is another notation for a partition due to Frobenius 
(see \cite{Mac}). \\
\subsection{Baker-Akhiezer functions and bilinear identities}
Vertex operators $V_\infty(z)$, $V_\infty^*(z)$ and 
$V_0(z)$, $V_0^*(z)$ act on the
space $C[t_1,t_2,\dots ]$ of polynomials in infinitely many variables, 
and are defined by the formulae:
\be\label{vertex}
V_\infty (z)=z^Me^{\xi({\bf t},z)}e^{-\xi(\tilde{\partial},z^{-1})},\quad
V^*_\infty (z)=z^{-M}e^{-\xi({\bf t},z)}e^{\xi(\tilde{\partial},z^{-1})} ,
\ee
\be
V_0(z)=z^{-M}e^{\xi({\bf t}^*,z^{-1})}e^{-\xi(\tilde{\partial}^*,z)},\quad
V^*_0(z)=z^Me^{-\xi({\bf t}^*,z^{-1})}e^{\xi(\tilde{\partial}^*,z)} ,
\ee
where $\tilde{\partial}=(\frac{\partial}{\partial t_1},
\frac{1}{2}\frac{\partial}
{\partial t_2},\frac{1}{3}\frac{\partial}{\partial t_3},\dots)$,
$\tilde{\partial}^*=(\frac{\partial}{\partial t_1^*},
\frac{1}{2}\frac{\partial}
{\partial t_2^*},\frac{1}{3}\frac{\partial}{\partial t_3^*},\dots)$. 

We have the rules of the bosonization:
\begin{eqnarray}
\l M+1|e^{H({\bf t})}\psi(z)=V_\infty (z)\l M|e^{H({\bf t})},\quad
 \l M-1|e^{H({\bf t})}\psi^*(z)=V^*_\infty (z)\l M |e^{H({\bf t})} ,
\end{eqnarray}
\begin{eqnarray}
\psi^*(z)e^{H^*({\bf t}^*)}|M\r=V^*_0(z) e^{H^*({\bf t}^*)}|M+1\r ,\quad
 \psi (z)e^{H^*({\bf t}^*)}|M\r=V_0(z)e^{H^*({\bf t}^*)}|M-1\r .
\end{eqnarray}

The Baker-Akhiezer functions and conjugated Baker-Akhiezer functions are:
\begin{eqnarray}\label{baker}
w_\infty (M,{\bf t},{\bf t}^*,z)=\frac{V_\infty(z)\tau}{\tau},\quad
w^*_\infty (M,{\bf t},{\bf t}^*,z)=\frac{V^*_\infty(z)\tau}{\tau} ,\\
w_0(M,{\bf t},{\bf t}^*,z)=\frac{V_0(z)\tau(M+1)}{\tau(M)},\quad
 w^*_0(M,{\bf t},{\bf t}^*,z)=\frac{V^*_0(z)\tau(M-1)}{\tau(M)} ,
\end{eqnarray}
where 
\begin{equation}\label{taufunction}
\tau(M,{\bf t},{\bf t}^*)=
\langle M|e^{H({\bf t})}ge^{H^*({\bf t}^*)}|M\rangle .
\end{equation}
Both KP and TL hierarchies are described by the bilinear identity:
\be\label{bi}
\oint w_\infty (M,{\bf t},{\bf t}^*,z)
w^*_\infty (M',{\bf t}',{{\bf t}'}^*,z)dz=
\oint w_0 (M,{\bf t},{\bf t}^*,z^{-1})
w^*_0 (M',{\bf t}',{{\bf t}'}^*,z^{-1})z^{-2}dz ,
\ee
which holds for any ${\bf t},{\bf t}^*,{\bf t}',{{\bf t}'}^* $
for any integers $M,M'$.\\

The Schur functions $s_{{\bf n}}({\bf t})$ are well-known examples of
tau-functions which correspond to rational solutions of the KP
hierarchy. It is known that not any linear combination of Schur 
functions turns to be a KP tau-function, in order to find these combinations
one should solve bilinear difference equation,  see \cite{JM},
which is actually a version of discrete Hirota equation.
Below we shall present KP tau-functions which are infinite series of Schur 
polynomials, and which turn to be known hypergeometric functions (\ref{vh}),(\ref{hZ}). We shall use the fermionic representation of tau-function
 \cite{JM}.

\section{ Hypergeometric functions related to Schur functions }
\subsection{KP tau-function $\tau_r(M,{\bf t},\beta )$}
Let $r$ be a function of one variable. 
Let $D=z\frac{d}{dz}$ acts on the basis $\{ z^n ;n \in Z \}$
of functions holomorphic in the  punctured disk $0<|z|<1$ . 
Then we put $r(D)z^n =r(n)z^n$. 
All functions of operator $D$ which we consider below are 
given via their eigenvalues on this basis. 

Let us consider an abelian subalgebra in ${\widehat {gl}}(\infty)$ formed
by the  set of fermionic operators
\begin{equation}\label{dopsim}
A_k=\frac{1}{2\pi i}
\oint \psi^*(z) \left(\frac{1}{z}r(D)\right)^k \psi(z),
\quad k=1,2,\dots  ,
\end{equation}
where the operator $r(D)$ acts on all functions of $z$ from the 
right hand side. In other terms
\begin{equation}\label{dopsim'}
A_k=
\sum_{n=-\infty}^\infty \psi^*_{n-k} \psi_n r(n)r(n-1)\cdots r(n-k+1),
\quad k=1,2,\dots  .
\end{equation}
We have $[A_m,A_k]=0$ for each $m,k$.
Fermionic operators (\ref{dopsim'}) resemble Toda lattice Hamiltonians
$-H^*_k$ (\ref{hamiltonians}), and coincide with them if $r(n)=1, n \in Z$.

For the collection of independent variables 
$\beta=(\beta_1,\beta_2,\dots)$ we denote
\begin{equation}\label{Abeta}
A(\beta)=\sum_{n=1}^\infty \beta_nA_n .
\end{equation}
For the partition ${\bf n}=(n_1,\dots ,n_k)$ and a function of one variable 
$r$, let us introduce the  notation
\begin{equation}\label{r_n}
r_{{\bf n}}(M)=\prod_{i=1}^{k}
r(1-i+M)r(2-i+M)\cdots r(n_i-i+M).
\end{equation}
We set $r_{{\bf 0}}(M)=1$.
Using the notation from (\ref{glemma}) we have\\
{\bf Lemma 2} The following formula holds
\be
\l 0|\psi^*_{i_1}\cdots\psi^*_{i_s}\psi_{-j_s}
\cdots\psi_{-j_1}e^{-A(\beta )}  |0\r
 =(-1)^{j_1+\cdots +j_s} 
r_{{\bf n}}(0)s_{{\bf n}}(\beta).
\ee
The proof is achieved by a direct calculation using 
$e^A=1+A+\frac 12 A^2+\cdots$, (\ref{dopsim'}), 
the hook decomposition of ${\bf n}$ and (\ref{glemma}).\\
Let us consider the  tau-function (\ref{taucorKP}) 
of the KP hierarchy 
\be\label{tauhyp1}
\tau_r(M,{\bf t},\beta ):=\langle M|e^{H({\bf t})}
e^{-A({\bf \beta })}  |M\rangle .
\ee
Using Taylor expanding $e^H=1+H+\cdots$ and {\em Lemma 1}, {\em Lemma 2} we 
easily get
\bprop 
We have the expansion:
\be\label{tauhyp}
\tau_r(M,{\bf t},\beta )
= \sum_{{\bf n}}
r_{ \bf n }(M)s_{\bf n }({\bf t})s_{\bf n }({\bf \beta }) .
\ee
\eprop
We shall not consider the problem of convergence of this series. 
The variables $M,{\bf t}$ play the role of KP higher
times, $\beta$ is a collection of group times for a commuting
subalgebra of additional symmetries of KP (see \cite{OW,D',ASM} and 
Remark 7 in \cite{O}). From different point of view (\ref{tauhyp}) 
is a tau-function of two-dimensional Toda lattice \cite{UT} with two sets of
continuous variables ${\bf t}$, $\beta$ and one discrete variable $M$. 
Formula (\ref{tauhyp}) is symmetric with respect to   
${\bf t} \leftrightarrow\beta$.
This 'duality' supplies us with the string equations \cite{T} which 
characterize a tau-function of hypergeometric type (see below).
In \cite{Tinit} the similar expansions to (\ref{tauhyp}) were considered,
without specifying the coefficients and in a different context.\\
For given $r$ we define the function $r'$:
\be
r'(n):=r(-n) .
\ee
\bprop 
We have the  involution:
\be\label{tausim}
\tau_{r'}(-M,-{\bf t},-\beta )
=\tau_r(M,{\bf t},\beta ).
\ee
\eprop
The proof follows from the relations
\be
r'_{\bf n}(M)=r_{{\bf n}'}(-M),\quad
s_{\bf n}({\bf t})=s_{{\bf n}'}(-{\bf t}).
\ee

Now let us introduce 
\begin{equation}\label{tdopsim}
\tilde{A}_k=-\frac{1}{2\pi i}
\oint \psi^*(z) \left(\tilde{r}(D)z\right)^k \psi(z),
\quad (k=1,2,\dots),\quad
\tilde{A}(\beta)=\sum_{n=1}^\infty \tilde{\beta}_n\tilde{A}_n .
\end{equation}
Then we have the following generalization of Proposition 1:

\bprop
\begin{equation}\label{trr}
\langle M|e^{\tilde{A}( {\tilde {\beta }})}
e^{-A({\bf \beta })}  |M\rangle
= 
\sum_{{\bf n}}
(\tilde{r}r)_{ \bf n }(M)s_{\bf n }(\tilde{\beta })s_{\bf n }({\bf \beta }).
\end{equation}

\eprop
\bremark
This expansion has the following interpretation. 
If one uses vacuum vectors $\l M| |M\r$ for normal ordering 
$:A:=A-\l M| A|M\r$ in the formula for $\widehat{gl}(\infty)$
commutation relation (\ref{commutator}),
he gets different $\widehat{gl}(\infty)$ 2-cocycles $c_M$ which are 
cohomological to $c_0$ (\ref{cocycle}).
The value of  $c_M$ on the elements $\tilde{A}_1,A_1$ is
$\tilde{r}(M)r(M)$:
\be
c_M(\tilde{A}_1,A_1)=\left(\tilde{r}r\right)(M).
\ee
Formula (\ref{trr}) is an expansion of $\widehat{GL}(\infty)$ group 2-cocycle,
evaluated on the elements $e^{\tilde{A}( {\tilde{\beta}})}$,$e^{-A(\beta )}$, 
in terms of $r(M)$, see also {\em Appendix ``Gauss factorization 
problem etc.''}\\
\eremark 

In what follows we put $\tilde{r}=1$, since (\ref{trr})
depends only on $\tilde{r}r$.

\subsection{$H_0({\bf T})$, twisted fermions $\psi({\bf T},z),     
\psi^*({\bf T},z)$ and bosonization rules}
Let $r\neq 0$, and put $r(n)=e^{T_{n-1}-T_n}$, where the variables 
$T_n$ are defined
up to a constant independent of $n$. We define a Hamiltonian 
$H_0({\bf T})\in \widehat{gl}(\infty)$  (all $T_n \in C$ are finite):
\be\label{H0+}
H_0({\bf T}):= \sum_{n=-\infty}^{\infty} T_n :\psi^*_n \psi_n: ,
\ee
which produces the transformation (\ref{gauge}): 
\be\label{prop41}
e^{\mp H_0({\bf T})}\psi_n e^{\pm H_0({\bf T})}=e^{\pm T_n}\psi_n,
\quad
e^{\mp H_0({\bf T})}\psi^*_n e^{\pm H_0({\bf T})}=e^{\mp T_n}\psi^*_n ,
\ee
\be\label{prop43}
e^{H_0({\bf T})}\tilde{A}(\tilde{{\beta}}) e^{-H_0({\bf T})}=
H(\tilde{{\beta}}),\quad
e^{-H_0({\bf T})}A({\beta}) e^{H_0({\bf T})}=-H^*({\beta}) .
\ee

Let $r\neq 0$. It is convenient to consider the  fermionic operators:
\begin{eqnarray}\label{kruchferm}
\psi({\bf T},z)=e^{H_0({\bf T})}\psi(z)e^{-H_0({\bf T})}
=\sum_{n=-\infty}^{n=+\infty} e^{-T_n}z^n\psi_n ,\\
\psi^*({\bf T},z)=e^{H_0({\bf T})}\psi^*(z)e^{-H_0({\bf T})}=
\sum_{n=-\infty}^{n=+\infty} 
e^{T_n}z^{-n}\psi^*_n\frac{dz}{z} .
\end{eqnarray}

For the variables ${\bf t}^{+}({\bf x}_{(N)})$ and 
${\bf t^*}^{+}({\bf y}_{(N)})$,  
and for the ``Hamiltonians'' $A$ and ${\tilde{A}}$ defined by (\ref{dopsim}),
(\ref{tdopsim}),
  one can derive the bosonization rules:
\begin{eqnarray}\label{mm*}
e^{-A({\bf t^*}^{+}({\bf y}_{(N)}))}|M\r=\frac{\psi({\bf T},y_1)\cdots\psi({\bf T},y_N)|M-N\r }{\Delta^{+}(M, N,{\bf T},{\bf y}_{(N)})} ,\\
e^{-A({\bf t^*}^{-}({\bf y}_{(N)}))}|M\r=\frac{\psi^*({\bf T},y_1)\cdots\psi^*({\bf T},y_N)|M+N\r}
{\Delta^{-}(M, N,{\bf T},{\bf y}_{(N)})} ,\\
\l M|e^{{\tilde{A}}({\bf t}^{+}({\bf x}_{(N)}))} =\frac{
\l M-N|\psi^*(-{\bf \tilde{T}},\frac{1}{x_N})\cdots\psi^*(-{\bf \tilde{T}},\frac{1}{x_1})}
{{\tilde \Delta}^{+}(M, N,{\bf \tilde{T}},{\bf x}_{(N)})},\\
\l M|e^{\tilde{A}({\bf t}^{-}({\bf x}_{(N)}))} =\frac{
\l M+N|\psi (-{\bf \tilde{T}},\frac{1}{x_N})\cdots\psi (-{\bf \tilde{T}},\frac{1}{x_1})}
{{\tilde \Delta}^{-}(M, N,{\bf \tilde{T}},{\bf x}_{(N)})} .
\end{eqnarray}
Here $\tilde{T}_n$ are related to $\tilde{A}$ via (\ref{tdopsim}) and 
$\tilde{r}(n)=e^{\tilde{T}_{n-1}-\tilde{T}_n}$.
Vandermond coefficients are
\begin{eqnarray}\label{vand}
\Delta^+(M, N,{\bf T},{\bf y}_{(N)})=
\frac{\prod_{i<j}(y_i-y_j)}{(y_1\cdots y_N)^{N-M}}
\frac{\tau(M,{\bf 0,T,0})}{\tau(M-N,{\bf 0,T,0})},\\
\Delta^{-}(M, N,{\bf T},{\bf y}_{(N)})=
\frac{\prod_{i<j}(y_i-y_j)}{(y_1\cdots y_N)^{M+N}}
\frac{\tau(M,{\bf 0,T,0})}{\tau(M+N,{\bf 0,T,0})},\\
{\tilde \Delta}^+(M, N,{\bf \tilde{T}},{\bf x}_{(N)})=
\frac{\prod_{i<j} (x_i-x_j)}{(x_1\cdots x_N)^{N-M-1}}
\frac{\tau(M,{\bf 0,\tilde{T},0})}{\tau(M-N,{\bf 0,\tilde{T},0})},\\
{\tilde \Delta}^{-}(M, N,{\bf \tilde{T}},{\bf x}_{(N)})=
\frac{\prod_{i<j} (x_i-x_j)}{(x_1\cdots x_N)^{N+M-1}}
\frac{\tau(M,{\bf 0,\tilde{T},0})}{\tau(M+N,{\bf 0,\tilde{T},0})}.
\end{eqnarray}
The notation $\tau(M,{\bf 0,T,0})$ is explained in the next Subsection,
see (\ref{H0+}),(\ref{H0-}).\\
Therefore in Miwa variables one can rewrite correlators (\ref{trr}):
\begin{eqnarray}\label{fermicor}
\l M|e^{{\tilde{A}}({\bf t}^{+}({\bf x}_{(N)}))}
e^{-A({\bf t^*}^{+}({\bf y}_{(N)}))}|M\r=\nonumber\\
\frac{
\l M-N|\psi^*(-{\bf \tilde{T}},\frac{1}{x_N})\dots
\psi^*(-{\bf \tilde{T}},\frac{1}{x_1})
\psi({\bf T},y_1)\cdots\psi({\bf T},y_N)|M-N\r}
{{\tilde \Delta}^{+}(M, N,{\bf \tilde{T}},{\bf x}_{(N)})
 \Delta^{+}(M, N,{\bf T},{\bf y}_{(N)})},
\end{eqnarray}
\begin{eqnarray}\label{fermicor-}
\l M|e^{\tilde{A}({\bf t}^{-}({\bf x}_{(N)}))}
e^{-A({\bf t^*}^{-}({\bf y}_{(N)}))}|M\r=\nonumber\\
\frac{
\l M+N|\psi (-{\bf \tilde{T}},\frac{1}{x_N})\cdots\psi (-{\bf \tilde{T}},\frac{1}{x_1})
\psi^*({\bf T},y_1)\cdots\psi^*({\bf T},y_N)|M+N\r}
{{\tilde \Delta}^{-}(M, N,{\bf \tilde{T}},{\bf x}_{(N)})
\Delta^{-}(M, N,{\bf T},{\bf y}_{(N)})}.  
\end{eqnarray}

\subsection{Toda lattice tau-function $\tau(M,{\bf t},{\bf T},{\bf t}^*)$ }

Now let us consider the  {\em Toda lattice} tau-function 
(\ref{taucor}), which depends on the three
sets of variables ${\bf t},{\bf T},{\bf t}^*$ and on $M \in Z$:
\begin{equation}\label{tau3}
\tau(M,{\bf t},{\bf T},{\bf t}^*)=
\langle M|e^{H({\bf t})}
\exp\left(\sum_{-\infty}^{\infty}T_n:\psi_n^*\psi_n:\right)
e^{H^*({\bf t}^*)}|M\rangle ,
\end{equation}
where $:\psi_n^*\psi_n:=\psi^*_n\psi_n -\l 0| \psi_n^*\psi_n |0\r $.
Since the operator $\sum_{-\infty}^{\infty}:\psi_n^*\psi_n:$ commutes with
all elements of the $\widehat{gl} (\infty)$ algebra, one can put $T_{-1}=0$
in (\ref{tau3}). With respect to the KP and the TL dynamics the 
times $T_n$ have a meaning of integrals of motion. With respect to
each pair of times $(t_m,T_n)$ one can consider the Liouville equation
related to (\ref{tau3}), see {\em Appendix 
``Equations with respect to ${\bf T}$ variables''}
 (the variables ${\bf t}^*$ plays 
the role of integrals of motion for these Liouville equations).
 
As we shall see the hypergeometric functions 
(\ref{ordinary}),(\ref{qordinary}),(\ref{vh}),(\ref{hZ1}) listed
in the Introduction are ratios of tau-functions (\ref{tau3})
evaluated at special values of times $M,{\bf t},{\bf T},{\bf t^*}$.
It is true only in the case when all parameters $a_k$ of the hypergeometric
functions are nonintegers. For the case when at least one of the indices  
$a_k$ is an integer, we will need a tau-function of an open Toda chain 
which will be considered in the next Sections.

Tau-function (\ref{tau3}) is linear in each $e^{T_n}$. It is
described by the  Proposition

\bprop
\be\label{tauhyp'}
\frac{\tau (M,{\bf t},{\bf T},{\bf t}^* )}
{\tau (M,{\bf 0},{\bf T},{\bf 0} )}
=1+ \sum_{{\bf n}\neq {\bf 0}}
e^{(T_{M-1}-T_{n_1+M-1})+(T_{M-2}-T_{n_2+M-2})+\cdots +
(T_{M-l}-T_{n_l+M-l}) }s_{\bf n }({\bf t})
s_{\bf n }({\bf  t }^*) .
\ee
The sum is going over all different partitions 
\be\label{part}
{\bf n}=(n_1,n_2,\dots,n_l),\quad l=1,2,3,... ,
\ee
excluding the partition ${\bf 0}$.\\
\eprop
Let $r \neq 0$,$\tilde{r} \neq 0$. Then we put
\be\label{rT}
r(n)=e^{T_{n-1}-T_{n}} ,\quad \tilde{r}(n)=
e^{\tilde{T}_{n-1}-\tilde{T}_{n}}.
\ee
Let us show the equivalence of (\ref{trr}) and (\ref{tauhyp'}) in this case.
We have $\tau (0,{\bf 0},{\bf T},{\bf 0} )=1$ and
\be\label{H0+}
\tau (n,{\bf 0},{\bf T},{\bf 0} )
=e^{-T_{n-1}-\cdots -T_1-T_0}, \quad n>0,
\ee
\be\label{H0-}
\tau (n,{\bf 0},{\bf T},{\bf 0} ) =
e^{T_{n}+\cdots +T_{-2}+T_{-1}}, \quad n<0 .
\ee
\bprop
Let $\tau(n,{\tilde {\beta}},{\bf T},\beta)$ is Toda lattice 
tau-function (\ref{tau3}) and $\tau_{\tilde{r}r}(n,\tilde{\beta},\beta)$
is defined by (\ref{trr}), where $\tilde{r},r,{\bf T},\tilde{\bf T}$
are related by (\ref{rT}), the functions $r,\tilde{r}$ have no zeroes 
at integer values of argument then
\be\label{teplitsGauss}
\frac{\tau(n,{\tilde {\beta}},\tilde{{\bf T}}+{\bf T},\beta)}
{\tau(n,{{\bf 0}},\tilde{{\bf T}}+{\bf T},{\bf 0})}=
\l n | e^{\tilde{A} ({\tilde{\beta}})} e^{-A(\beta)} | n \r =
\tau _{\tilde{r}r}(n,\tilde{\beta},\beta).
\ee    
This proposition follows from formulas (\ref{prop41})-(\ref{prop43}).
\eprop

For $\tilde{r}=1$ we can put $\tilde{\beta}={\bf t}$. Then
the next equations hold
\begin{eqnarray}\label{rToda}
\partial_{t_1}\partial_{\beta_1}\phi_n=
r(n)e^{\phi_{n-1}-\phi_{n}}-
r(n+1)e^{\phi_{n}-\phi_{n+1}} ,\quad
e^{-\phi_n}=\frac{\tau_r(n+1,{\bf t},\beta)}{\tau_r(n,{\bf t},\beta)}
,\\
\left(\tau(n):=\tau_r(n,{\bf t},\beta)\right) \qquad \tau(n)\partial_{\beta_1}
\partial_{t_1}\tau(n)-
\partial_{t_1}\tau(n)
\partial_{\beta_1}\tau(n)=r(n)\tau(n-1)\tau(n+1).\label{hirota}
\end{eqnarray}
As we shall see eqs. (\ref{rToda}) and (\ref{hirota}) are still true in case
$r(n)$ has zeroes.\\
If the function $r$ has no integer zeroes, using the change of variables
\be
\varphi_n=-\phi_n - T_n ,
\ee
we obtain Toda lattice equation in standard form \cite{UT}:
\be\label{Toda}
\partial_{t_1}\partial_{t^*_1}\varphi_n=e^{\varphi_{n+1}-\varphi_{n}}-
e^{\varphi_{n}-\varphi_{n-1}}.
\ee
As we see the variables $T_n$ might have the meaning of asymptotic values of
the fields $\phi_n$ for the class of tau-functions (\ref{tau3})
which is characterized by the property $\varphi_n \to 0$ as $t_1 \to 0$.  

\subsection{Toda lattice consisted of open parts}

Now we consider tau-function (\ref{tau3}) with the modification of the 
definition of flows.

Definition. Let us introduce the function $\delta$ which is equal to
zero when $r$ is equal to zero and is equal to unity otherwise:
\be
\delta(n)=0\quad if\quad r(n)=0,\qquad \delta(n)=1\quad if\quad r(n) \neq 0 .
\ee
Given collection of zeroes ${\bf m}$ of $r$:
\be\label{integralzeroes}
{\bf m}=\{M_i \in Z\},\quad M_{i+1}>M_i,\quad r(M_i)=0.
\ee
we construct Hamiltonians labeled by  ${\bf m}$:
\begin{eqnarray}
H_{-k}({\bf m})=\sum_{n=-\infty}^{+\infty}\delta(n)
\delta(n-1)\cdots \delta(n-k+1)
\psi_n\psi^*_{n-k},\quad
H^*({\bf m};\beta)=\sum H_{-k}({\bf m})\beta_k .\quad
\\
H_{k}({\bf m})=\sum_{n=-\infty}^{+\infty}\delta(n+1)
\delta(n+2)\cdots \delta(n+k)
\psi_n\psi^*_{n+k},\qquad
H({\bf m};{\bf t})=\sum H_{k}({\bf m})t_k .\quad
\end{eqnarray}
The tau-function of the open TL we are interested in, see 
(\ref{deltaToda}) below, can be written in the  three equivalent 
forms:
\begin{eqnarray}
\label{taucoropen}
\tau_{op}({\bf m};M,{\bf t},\beta)=
\langle M|e^{H({\bf t})}
\exp\left(\sum_{-\infty}^{\infty}T_n:\psi_n^*\psi_n:\right)
e^{H^*({\bf m};\beta)}|M \rangle =\qquad \qquad \qquad \\
\langle M|e^{H({\bf m};{\bf t})}
\exp\left(\sum_{-\infty}^{\infty}T_n:\psi_n^*\psi_n:\right)
e^{H(\beta)}|M \rangle =
\langle M|e^{H({\bf m};{\bf t})}
\exp\left(\sum_{-\infty}^{\infty}T_n:\psi_n^*\psi_n:\right)
e^{H({\bf m};\beta)}|M \rangle, \quad
\end{eqnarray}
where $T_n \in C$ are some constants (times).
This tau-function has the property:
\be
 M_i \in {\bf m} \Rightarrow
\tau_{op}({\bf m};M_i,{\bf t},\beta)=1
\ee
for all values of times ${\bf t},\beta$. 
If one consider
\be
\varphi_n=\log \frac{\tau_{op}({\bf m};n+1,{\bf t},{\bf T},\beta)}
{\tau_{op}({\bf m};n,{\bf t},{\bf T},\beta)}
\ee
he comes to the equation of an open TL:
\be\label{deltaToda}
\partial_{t_1}\partial_{\beta_1}\varphi_n=
\delta(n)e^{\varphi_{n-1}-\varphi_{n}}-
\delta(n+1)e^{\varphi_{n}-\varphi_{n+1}} . 
\ee
The set of fields $\varphi_n$ solves a number of open lattice problems 
in the set of intervals:
\begin{equation}\label{semiinfh'} 
\{\varphi_n , \quad n < M_1  \},
\end{equation}
\begin{equation}\label{finitechainh'}
\{\varphi_n , \quad M_{i}\leq n < M_{i+1},\quad  M_{i+1}- M_{i}>1 \}, 
\end{equation}
\begin{equation}\label{semiinf'h'}
\{\varphi_n , \quad  n > M_s \}.
\end{equation}
The tau-function describes a set of open Toda lattices between each pair of
neighbor zeroes (between neighbor zeroes  $M_{i+1}$,$M_i$
there is an open chain with $M_{i+1}-M_i$ number of sites), and
two semiinfinite Toda lattices, one of them ends on the smallest
zero and the other on the largest zero.

\subsection{Properties of the tau function $\tau_r$ when function r(n) 
has zeroes} 

Now let us introduce a set of $T_n$ variables with the help
of relations
\be\label{rTo}
r(n)=e^{T_{n-1}-T_n}
\ee
for all $n$ where $ r(n)\neq 0$.
Equation (\ref{rTo}) define variables $T_n$ uniquely 
 up to an integration constant in each of the 
intervals (the number of the constants is equal to the number of intervals)
\begin{equation}\label{semiinfh} 
\{T_n,\quad n < M_1 \},
\end{equation}
\begin{equation}\label{finitechainh}
\{T_n,\quad M_{i}\leq n < M_{i+1},\quad  M_{i+1}- M_{i}>1 \}, 
\end{equation}
\begin{equation}\label{semiinf'h}
\{T_n,\quad M_s \leq n \}
\end{equation}
separately. In case of there are zeroes such that  $M_{i+1}=M_i+1$
one can define variables $T_{M_i}$, however we will not need them.

We introduce the Hamiltonian
\be
H_0({\bf T})=\sum_n T_n :\psi^*_n\psi_n: ,
\ee
where sum is over all $n$ satisfying  one of the equations (\ref{semiinfh}),
(\ref{finitechainh}) or (\ref{semiinf'h}).\\

We have
\be
A_k=-e^{H_0({\bf T})}H_{-k}({\bf m})e^{-H_0({\bf T})},\quad
A(\beta)= -e^{H_0({\bf T})}H^*({\bf m},\beta)e^{-H_0({\bf T})}.
\ee

\bprop
Let $\tau_{op}({\bf m},n,{\bf t},{\bf T},\beta)$ is Toda lattice 
tau-function (\ref{taucoropen}), and $\tau_{r}(n,{\bf t},\beta)$
is defined by (\ref{tauhyp1}), where $r$ and ${\bf T}$
are related by (\ref{rTo}), the functions $r$ has zeroes 
at integer values of argument described by (\ref{integralzeroes})
and for $r\neq 0$ the set of variables ${\bf T}$ is related to
$r$ by (\ref{rTo})
\be\label{teplitsGausso}
\frac{\tau_{op}({\bf m},n,{\bf t},{\bf T},\beta)}
{\tau_{op}({\bf m},n,{{\bf 0}},{\bf T},{\bf 0})}=
\l n | e^{H ({\bf t})} e^{-A(\beta)} | n \r =
\tau _{r}(n,{\bf t},\beta).
\ee  
Equations (\ref{rToda}) and (\ref{hirota}) are still true in case
$r(n)$ has zeroes.
\eprop
Hirota equation (\ref{hirota}) can be viewed as recurrent relation 
which expresses tau-function with discrete Toda lattice variable $n$ via 
$\tau_r(M_i\pm 1,{\bf t},\beta),\tau_r(M_i,{\bf t},\beta)=1$.
It follows from (\ref{r_n}),(\ref{tauhyp}) that
\be
r(M_k)=0 \Rightarrow \tau_r (M_k,{\bf t},\beta)=1.
\ee
Then from (\ref{r_n}),(\ref{tauhyp}) we see the following. In the region 
(\ref{finitechainh'})
the series (\ref{tauhyp}) has only a finite number of nonvanishing terms.
For the region (\ref{semiinf'h'})
the sum is only over the Young diagrams ${\bf n}$ of the length 
$l({\bf n})<M-M_1$. For the region
(\ref{semiinfh'})
only those diagrams ${\bf n}$ for which the conjugated diagrams 
{${\bf n'}$ have length  $l({\bf n'})\le M_s-M$ contribute the 
series (\ref{tauhyp}). 

In {\em Appendix} we shall write down a system of orthogonal 
polynomials related to ${\bf m}$.

\bremark
There are two different ways to restrict the sum 
(\ref{tauhyp}) to a sum over partitions of length $l({\bf n})\leq N$
(or over $l({\bf n'})\leq N$).
The second way is to use so-called Miwa's change of variables.
\eremark

\subsection{Notations}
In order to simplify notations, we shall omit 
additional argument ${\bf m}$ and subindex which distinguish 
TL tau-function (\ref{tau3}) and open TL tau-functions (\ref{taucoropen}).
Instead of $\tau_{op}({\bf m},n,{\bf t},{\bf T},{\bf t}^*)$ we shall write
$\tau(n,{\bf t},{\bf T},{\bf t}^*)$.
The notation $\tau_r(M,{\bf t},\beta)$ will be used only for the 
KP tau-function (\ref{tauhyp1}). Also $\beta = {\bf t}^*$.
When TL higher times are expressed via Miwa change (\ref{Miwa+}) or
 (\ref{Miwa-}), sometimes we shall put the argument ${\bf x}_{(N)}$ at the 
place of the argument ${\bf t}$ and the argument ${\bf y}_{(N)}$ at the 
place of ${\bf t}^*$, for instance $\tau_r(M,{\bf x}_{(N)},{\bf t}^*)$, 
$\tau_r(M,{\bf t},{\bf y}_{(N)})$,
 $\tau_r(M,{\bf x}_{(N)},{\bf y}_{(N)})$.

\subsection{Linear equations for the tau-function $\tau_r$}

Here we shall write down linear equations, which follow from the 
explicit fermionic representation of the tau-function (\ref{tauhyp1})
via the bosonization formulae (\ref{fermicor}) and .(\ref{fermicor-})
These equations may be also viewed as the constraints which result in the 
string equations.
For the variables ${\bf t}^{-}({\bf x}_{(N)})$, using $\l M|A=0$ and making profit of the 
relation $A_k=e^{H_0}H_{-k}e^{-H_0}$ inside the fermionic correlator 
(\ref{fermicor}), we get the 
partial differential equations for the tau-function (\ref{tauhyp}):
\be\label{linur}
\frac{\partial \tau_r(M, {\bf t}^{-}({\bf x}_{(N)}), {\bf t}^*)}{\partial t^*_k}=
\frac{1}{{\tilde \Delta}}\left(\sum_{i=1}^N (x_ir(-D_{x_i}))^k\right)
{\tilde \Delta}
\tau_r(M, {\bf t}^{-}({\bf x}_{(N)}), {\bf t}^*),
\ee
where 
${\tilde \Delta}={\tilde \Delta}^{-}(M, N,{\bf 0},{\bf x}_{(N)})$.
These equations have the meaning
of string constraint equations for the tau-function (\ref{tauhyp}).
In variables ${\bf t^*}^{-}({\bf y}_{(\infty)})$ we can rewrite (\ref{linur}):
\begin{eqnarray}
(-1)^{k}\sum_{i=1}^{+\infty}\frac{e_{k-1}\left(\frac{1}{y_1},\dots,
\frac{1}{y_{i-1}},\frac{1}{y_{i+1}},\dots\right)}{\prod_{j\neq i}(1-\frac{y_i}{y_j})}\frac{\partial \tau_r(M,{\bf t}^{-}({\bf x}_{(N)}), {\bf t^*}^{-}({\bf y}_{(\infty)}))}
{\partial y_i}=\nonumber\\
\frac{1}{{\tilde \Delta}}\left(\sum_{i=1}^N (x_ir(-D_{x_i}))^k\right)
{\tilde \Delta}
\tau_r(M, {\bf t}^{-}({\bf x}_{(N)}), {\bf t^*}^{-}({\bf y}_{(\infty)})),
\end{eqnarray}
where $e_k({\bf y})$ is a symmetric function defined through the relation
$\prod_{i=1}^{+\infty}(1+ty_i)=\sum_{k=0}^{+\infty}t^ke_k({\bf y})$.
Also we have
\begin{eqnarray}\label{linur2}
 \left(\sum_{k=1}^{M+N-1} k -\sum_{i=1}^N D_{x_i}\right)
{\tilde \Delta}^{-}\tau(M, {\bf t}^{-}({\bf x}_{(N)}),{\bf T},{\bf t}^{*-}({\bf y}_{(N')}))\Delta^{-}=\nonumber\\
 \left(\sum_{k=1}^{M+N'-1} k -\sum_{i=1}^{N'}\left(\frac{1}{y_i}D_{y_i}y_i\right)\right)
{\tilde \Delta}^{-}\tau(M, {\bf t}^{-}({\bf x}_{(N)}),{\bf T},{\bf t}^{*-}({\bf y}_{(N')}))\Delta^{-},
\end{eqnarray}
where ${\tilde \Delta}^{-}={\tilde \Delta}^{-}(M, N,{\bf 0},{\bf x}_{(N)})$ and
$\Delta^{-}=\Delta^{-}(M, N,{\bf 0},{\bf y}_{(N')})$.
This formula is obtained by the insertion of the fermionic operator
$ res_z:\psi^*(z)z\frac{d}{dz}\psi(z):$ inside the fermionic correlator.
These formulae can be also written in terms of higher KP and TL times, 
with the help of vertex operator action,
see the {\em Appendix ``Vertex operator action''}.
Then the relation (\ref{linur}) is the infinitesimal version of 
(\ref{stringvert1}), while the relation (\ref{linur2}) is the 
infinitesimal version of (\ref{stringvert2}).

\subsection{Determinant formulae}
With the help of Wick theorem \cite{JM} one obtains the 
formulae.
\bprop{A generalization of Milne's determinant formula}
\begin{equation}\label{det1}
\tau_r(M,{\bf t}^{+}({\bf x}_{(N)}),{\beta})=\frac{
\det\left(x_i^{N-k}\tau_r(M-k+1,{\bf t}^{+}(x_i),{\beta}) \right)_{i,k=1}^{N}}{
\det\left(x_i^{N-k}\right)_{i,k=1}^N} .
\end{equation}
{\bf Proof}\\
\begin{eqnarray}
\tau_r(M,{\bf t}^{+}({\bf x}_{(N)}),{\beta})=\l M|e^{H({\bf t}^{+}({\bf x}_{(N)})}e^{-A({\beta})}|M\r=\\
\label{miwatau}
\frac{x_1^{N-M-1}\cdots x_N^{N-M-1}}{\prod_{i<j} (x_i-x_j)}\l M|\psi_{M-1}
\cdots \psi_{M-N}\psi^*(\frac{1}{x_N})\cdots\psi^*(\frac{1}{x_1})
e^{-A({\beta})}|M\r=\\
\frac{(x_1\cdots x_N)^{N-M-1}}{\prod_{i<j} (x_i-x_j)}
\det\left(\l M|\psi_{M-k}\psi^*(\frac{1}{x_i})e^{-A({\beta})}|M\r\right)_{i,k=1}^N=\\ \frac{
\det\left(x_i^{N-k}\tau_r(M-k+1,{\bf t}^{+}(x_i),{\beta}) \right)_{i,k=1}^{N}}{
\det\left(x_i^{N-k}\right)_{i,k=1}^N} 
\end{eqnarray}
Last equality follows from:
\begin{eqnarray}
\l M|\psi_{M-k}\psi^*(\frac{1}{x_i})e^{-A({\beta})}|M\r=\\
=\l M|\psi_{M-1}\cdots \psi_{M-k+1}\psi_{M-k}\psi^*(\frac{1}{x_i})e^{-A({\beta})}\psi^*_{M-k+1}\cdots \psi^*_{M-1}|M\r+\\
+\sum_{j=1}^{k-1} a^k_j(\beta)\l M|\psi_{M-1}\cdots \psi_{M-k+j}\psi^*(\frac{1}{x_i})e^{-A({\beta})}\psi^*_{M-k+j+1}\cdots \psi^*_{M-1}|M\r=\\
=\l M-k+1|\psi_{M-k}\psi^*(\frac{1}{x_i})e^{-A({\beta})}|M-k+1\r+\\
+\sum_{j=1}^{k-1} a^k_j(\beta)\l M-k+1+j|
\psi_{M-k+j}\psi^*(\frac{1}{x_i})e^{-A({\beta})}|M-k+1+j\r=\\
=x_i^{M-k+1}\tau_r(M-k+1,{\bf t}^{+}(x_i),\beta)+\sum_{j=1}^{k-1} a^k_j(\beta) x_i^{M-k+1+j} \tau_r(M-k+1+j, {\bf t}^{+}(x_i), \beta)
\end{eqnarray}
Where the functions $a^k_j(\beta)$ must be derived as the results of action of operator $e^{-A(\beta)}$ on the fermions $\psi_{M-1},\dots , \psi_{M-k}$. Thus we have:
\begin{eqnarray}
x_i^{N-M-1}\l M|\psi_{M-k}\psi^*(\frac{1}{x_i})e^{-A({\beta})}|M\r=\\
=x_i^{N-k}\tau_r(M-k+1,{\bf t}^{+}(x_i),\beta)+\sum_{l=1}^{k-1} a^k_{k-l}(\beta) x_i^{N-l} 
\tau_r(M-l+1,{\bf t}^{+}(x_i), \beta)
\end{eqnarray}
\eprop
\bprop
For $r\neq 0$  we take a tau function $\tau_r(M,{\bf t}^+({\bf x}_{(N)}),{\bf t^*}^+({\bf y}_{(N)}))$ and  apply Wick's theorem.  We get the determinant formula:
\be\label{det2}
\tau_r(M,{\bf t}^+({\bf x}_{(N)}),{\bf t^*}^+({\bf y}_{(N)}))=\frac{\det \left(F(x_iy_j)\right)_{i,j=1}^N}
{{\tilde\Delta}^{+}(M, N,{\bf 0}, {\bf x}_{(N)})\Delta^{+}(M, N,{\bf T}, {\bf y}_{(N)})},
\ee
\be
F(x_iy_j)=
\l M-N |\psi^*\left(\frac{1}{x_i}\right) \psi\left({\bf T},y_j\right) |M-N \r.
\ee
\eprop

\subsection{Integral representations}
For the fermions (\ref{kruchferm}) we easily get the  relations:
\be
\int \psi({\bf T},\alpha z)d\mu(\alpha)=
\psi\left({\bf T}+{\bf T}(\mu),z\right),\quad
\int \psi^*\left(-{\bf T},\frac {1}{\alpha z}\right)d{\tilde \mu}(\alpha)=
\psi^*\left(-{\bf T}-{\bf T}({\tilde \mu}),\frac 1z \right)
\ee
where $\mu,{\tilde \mu}$ are some integration measures, and shifts
of times $T_n$ are defined in terms of the moments:
\be
\int \alpha^n  d\mu(\alpha)=e^{-T_n(\mu)},\quad
\int {\alpha}^n  d{\tilde \mu}(\alpha)=
e^{-T_n({\tilde \mu})}.
\ee
Therefore thanks to the bosonization 
formulae (\ref{fermicor}) we have the  
relations for the tau-function (below ${\bf t}^*$ is defined via
 (\ref{Miwa+}))
\bprop{Integral representation formula holds}
\begin{eqnarray}\label{intrepr}
\int{\tilde \Delta}_{\bf \tilde{T}}({\tilde \alpha}{\bf x}_{(N)}) 
\frac{\tau\left(M,{\bf t}^{+}(\tilde{\alpha}{\bf x}_{(N)}),
{\bf T+\tilde{T}},{\bf t^*}^{+}({\bf \alpha y}_{(N)})\right)}
{\tau\left(M,{\bf 0},
{\bf T+\tilde{T}},{\bf 0}\right)}
\Delta_{\bf T}({\bf \alpha y}_{(N)})
\prod_{i=1}^N d{\tilde \mu}({\tilde \alpha}_i)\prod_{i=1}^N d\mu(\alpha_i)
\nonumber\\
={\tilde \Delta}_{\bf \tilde{T}+{\bf \tilde{T}(\tilde{\mu})}}({\bf x}_{(N)})
\frac{\tau\left(M,{\bf t}^{+}({\bf x}_{(N)}),{\bf T}+{\bf \tilde{T}}+
{{\bf \tilde{T}}}({\tilde \mu})+{\bf T}(\mu),{\bf t^*}^{+}({\bf y}_{(N)})
\right)}{\tau\left(M,{\bf 0},{\bf T}+{\bf \tilde{T}}+
{{\bf \tilde{T}}}({\tilde \mu})+{\bf T}(\mu),{\bf 0}\right)}
\Delta_{{\bf T}+{\bf T}(\mu)}({\bf y}_{(N)}).
\end{eqnarray}
where $\Delta_{{\bf T}}({\bf \alpha y}_{(N)})=
\Delta^+ (M, N, {\bf T}, {\bf \alpha y}_{(N)})$,
${\tilde \Delta}_{{\bf \tilde{T}}}({\tilde \alpha}{\bf x}_{(N)})=
{\tilde \Delta}^+(M, N, {\bf \tilde{T}},{\tilde \alpha}{\bf x}_{(N)})$,\\
${\bf \alpha y}_{(N)} =\left(\alpha_1 y_1,\alpha_2 y_2,\dots,
\alpha_N y_N\right)$ and $\tilde {\bf \alpha}{\bf x}_{(N)} =\left({\tilde \alpha}_1 x_1,
{\tilde \alpha}_2 x_2,\dots,{\tilde \alpha}_N x_N\right)$.
In particular
\begin{eqnarray}\label{leftintegral}
\int \frac{\tau\left(M,{\bf t},{\bf T},{\bf t^*}^{+}({\bf \alpha y}_{(N)})\right)}
{\tau\left(M,{\bf 0},{\bf T},{\bf 0}\right)}
\Delta_{{\bf T}}({\bf \alpha y}_{(N)})
\prod_{i=1}^N d\mu(\alpha_i)=\nonumber\\
\frac{\tau\left(M,{\bf t},{\bf T}+{\bf T}(\mu),{\bf t^*}^{+}({\bf y}_{(N)})\right)}
{\tau\left(M,{\bf 0},{\bf T}+{\bf T}(\mu),{\bf 0}\right)}
\Delta_{{\bf T}+{\bf T}(\mu)}({\bf y}_{(N)}).
\end{eqnarray}
\eprop
Remember that arbitrary linear combination of tau-functions
is not a tau-function.
Formulae (\ref{intrepr}) and also (\ref{leftintegral}) give
the integral representations for the tau-function (\ref{tauhyp1}).
It may help to express a tau-function with the help 
of a more simple one. 
If we choose the  integration measures:
\be\label{Hankel}
\frac{i}{2\pi}\int_{C}\psi({\bf T},\alpha z) e^{-\alpha}({-\alpha})^{-b-1}
d\alpha = \psi({\bf T}+{\bf T}^b,z),
\ee
\be\label{Gamma}
\int_0^\infty \psi({\bf T},\alpha z)
 e^{-\alpha}{\alpha}^{a}d\alpha
=\psi({\bf T}+{\bf T}^a,z),
\ee
\be\label{Beta}
\int_0^1\psi({\bf T},\alpha z) {\alpha}^{a}(1-\alpha )^{b-a-1}
d\alpha = \psi({\bf T}+{\bf T}^c,z),
\ee
where $C$ starts at $+\infty$ on the real axis, circles the origin in the
counterclockwise direction and returns to the starting point. Then
\be\label{Tabc}
T_n^b=\ln \Gamma(b+n+1), \quad T_n^a=-\ln \Gamma(a+n+1),\quad
T_n^c=\ln \frac {\Gamma(b+n+1)}{\Gamma(a+n+1)\Gamma(b-a)}.
\ee

Also consider the  $q$-integrals \cite{KV}:
\be\label{qHankel}
q^{-(a+n)(a+n+1)}\int_0^\infty \psi({\bf T},\alpha (1-q)z) E_q(-\alpha)
\alpha ^{a} d_q\alpha
=\psi\left({\bf T}+{\bf T}(a,q),z\right),
\ee
\be\label{qBeta}
\frac {1}{\Gamma_q(b-a)}
\int_0^1\psi({\bf T},\alpha z) {\alpha}^{a}\frac{(\alpha q;q)_\infty }
{(\alpha q^{b-a};q)_\infty }
d_q\alpha = \psi({\bf T}+{\bf T}(a,b,q),z).
\ee
Then
\be
T_n(a,q)=\ln \frac{1}{(1-q)^{n}\Gamma_q(a+n+1)},\quad T_n(a,b,q)=
\ln  \frac {\Gamma_q(b+n+1)}{\Gamma_q(a+n+1)}.
\ee
In the same way one can consider Miwa change (\ref{Miwa-}).
In the {\em Examples} below we shall present hypergeometric functions
listed in the {\em Subsections 1.2 and 1.3} as tau-functions of the type
(\ref{fermicor}).
Then we are able to write down integration formulae, namely 
(\ref{leftintegral}), which express ${}_{p+1}\Phi_{s}$ and  
${}_{p+1}\Phi_{s+1}$ in terms of  ${}_p\Phi_s$ with the help of
 (\ref{qHankel}), (\ref{qBeta}) and (\ref{intrepr}).
In \cite{Milne} different integral representation formula was presented,
which was based on the $q$-analog of Selberg's integral of Askey and
Kadell.
By taking the limit $q \to 1$ one can
consider functions  ${}_{p}F_s$. 
Using (\ref{Tabc}), one can express
${}_{p+1}{\mathcal{F}}_{s}$, ${}_{p+1}{\mathcal{F}}_{s+1}$ and 
${}_{p}{\mathcal{F}}_{s+1}$ 
 as integrals of
${}_p{\mathcal{F}}_s$ with the help of (\ref{Gamma}), (\ref{Beta}) and
(\ref{Hankel}) respectively. 

\subsection{Examples}
The main point of the paper
is the observation that if  $r(D)$ is a rational function
of $D$ then $\tau_r$ is a hypergeometric series. If  $r(D)$ is
a rational function of $q^D$ we obtain $q$-deformed hypergeometric series.
Now let us consider various $r(D)$.\\

{\bf Example 1}
Let $r=1$. One can put ${\bf T}=0$. Then one gets
\begin{equation}\label{ex1}
\tau_{r=1}(M,{\bf t},{\bf t}^* )=
\exp \left(\sum_{n=1}^\infty nt_nt^*_n\right) , 
\end{equation}
which is vacuum tau-function for the two-dimensional Toda lattice.
Formula (\ref{ex1}) is a manifestation of summation formulas for
Schur functions \cite{KV}. Let us note that this is also an example of
function ${}_1{\mathcal{F}}_0$(\ref{hZ1}).\\

{\bf Example 2}
Let $r(n)=n$, that is
$T_n=\ln \frac{1}{n!},n\geq 0$ and
$T_n=\ln (-1)^n(-n-1)!,n< 0$. Also let us put ${\bf t}^*=(t^*_1,0,0,...)$. 
 For $M=0,\pm1$ we get 
\be\label{1F0}
\tau_r(0,{\bf t},t^*_1)=1, \quad 
\tau_r(1,{\bf t},t^*_1)=
e^{\xi({\bf t},t_1^*)},\quad
\tau_r(-1,{\bf t},t^*_1)
=e^{-\xi({\bf t},t^*_1)}.
\ee
Here $t^*_1$ plays the role of spectral parameter for the vacuum
Baker-Akhiezer function. 
This fact is in accordance to the meaning of $t^*_1$ as a group time for 
the 
Galilean transformation \cite{O}.\\
Similar answers $\tau=e^{\pm \xi({\bf t},z^*)}$ one obtains if he 
substitutes 
$t^*_n=\pm n^{-1}(z^*)^n$ to (\ref{ex1}).
Let us note that (\ref{1F0}) are the functions ${}_1F_0(0;t_1,t_2,\dots )$,  ${}_1F_0(1;t_1,t_2,\dots )$ and  ${}_1F_0(-1;t_1,t_2,\dots )$ (\ref{hZ})
which will be described below in the {\em Example 3} (\ref{beskshur}).\\
{\bf Example 3} Let all parameters $b_k$ be nonintegers.
\be
{}_pr_s(D)=\frac{(D+a_1)(D+a_2)\cdots 
(D+a_p)}{(D+b_1)(D+b_2)\cdots (D+b_s)} . \label{op1}
\ee
If all $a_k$ are also nonintegers the
relevant ${\bf T}$ is:
\be
{}^pT^s_n=-\ln \frac{\Gamma(n+a_1+1)\Gamma(n+a_2+1)\cdots 
\Gamma(n+a_p+1)}{\Gamma(n+b_1+1)\Gamma(n+b_2+1)\cdots 
\Gamma(n+b_s+1)} . \label{op2}
\ee
For the correlator (\ref{tauhyp}) we have:
\be
\frac{{}^p\tau^s(M,{\bf t},{\bf T},{\bf t}^*)}
{{}^p\tau^s(M,{\bf 0},{\bf T},{\bf 0})}={}^p\tau_r^s(M,{\bf t},{\bf t}^*)=
\sum_{{\bf n}}s_{{\bf n}}({\bf t})s_{{\bf n}}({\bf t}^*)
\frac{(a_1+M)_{{\bf n}}\cdots (a_p+M)_{{\bf n}}}
{(b_1+M)_{{\bf n}}\cdots (b_s+M)_{{\bf n}}} . \label{tau}
\ee

 If in formula (\ref{tau}) we put 
\be\label{t^*}
t^*_1=1 , \qquad t^*_i=0,\quad i>1 
\ee
then $s_{{\bf n}}({\bf t}^*)=H^{-1}_{{\bf n}}$, and 
we obtain the hypergeometric function 
related to Schur functions \cite{KV} (see \cite{Mac} for help):
\begin{eqnarray}\label{beskshur}
\frac{{}^p\tau^s(M,{\bf t},{\bf T},{\bf t}^*)}
{{}^p\tau^s(M,{\bf 0},{\bf T},{\bf 0})}=
{}^p\tau_r^s(M,{\bf t},{\bf t}^*)
={}_pF_s\left.\left(a_1+M,
\dots ,a_p+M\atop b_1+M,\dots ,b_s+M\right|t_1,t_2,\dots\right)=
\nonumber\\
\sum_{{\bf n}}\frac{(a_1+M)_{{\bf n}}\cdots (a_p+M)_{{\bf n}}}
{(b_1+M)_{{\bf n}}\cdots (b_s+M)_{{\bf n}}}
\frac{s_{{\bf n}}({\bf t})}{H_{{\bf n}}}.
\end{eqnarray}
In the last formula $H_{{\bf n}}$ is the following hook polynomial
(compare with (\ref{hp})):
\begin{eqnarray}
H_{{\bf n}}=\prod_{(i,j)\in {\bf n}} h_{ij},  \qquad h_{ij}=(n_i+n'_j-i-j+1).
\end{eqnarray}
We obtain ordinary hypergeometric function of one variable of type 
\be\label{onevar}
{}_{p-1}F_s(a_2,\dots,a_p;b_1,\dots,b_s;\pm t_1t^*_1)=
\tau_r(\pm 1,{\bf t},{\bf T},{\bf t}^* ) ,
\ee
 if we take $a_1=0$, ${\bf t}=(t_1,0,0,\dots)$, 
${\bf t}^* =(t^*_1,0,0,\dots)$ \cite{Sc}.
\\

For variables ${\bf t}^{+}({\bf x}_{(N)})$
the formula (\ref{beskshur}) turns out to be
\begin{eqnarray}\label{tauzon}
{}^p\tau_r^s(M, {\bf t}^{+}({\bf x}_{(N)}), {\bf t^*})=
{}_pF_s\left.\left(a_1+M,\dots ,a_p+M\atop b_1+M,\dots ,
b_s+M\right| {\bf x}_{(N)}\right)=
\nonumber\\
\sum_{{\bf n}\atop l({\bf n})\le N}\frac{(a_1+M)_{{\bf n}}\cdots (a_p+M)_{{\bf n}}}
{(b_1+M)_{{\bf n}}\cdots (b_s+M)_{{\bf n}}}
\frac{s_{{\bf n}}({\bf x}_{(N)})}{H_{{\bf n}}}.\qquad \qquad \qquad \qquad 
\qquad 
\end{eqnarray}
We got the hypergeometric function (\ref{hZ}) related to zonal
 polynomials for
the symmetric space $GL(N,C)/U(N)$ \cite{V}.
Here $x_i=z_i^{-1},i=1,...,N$ are the eigenvalues of the matrix 
 ${\bf X}$, and for zonal spherical polynomials there is the following 
matrix integral representation
\begin{equation}\label{zonmatint}
Z_{\bf n}({\bf X})=
Z_{\bf n}({\bf I}_N)\int_{U(N,C)}\Delta^{\bf n}
\left( U^*{\bf X}U\right)d_*U ,
\end{equation}
where $\Delta^{\bf n}\left({\bf X}\right)=\Delta^{n_1-n_2}_1
\Delta^{n_2-n_3}_2\cdots \Delta^{n_N}_N$ and
$\Delta_1,\dots\Delta_N$ are main minors of the matrix ${\bf X}$,
$d_*U$ is the invariant measure on $U(N,C)$, 
see \cite{V} for the details.\\
Taking $N=1$ we obtain the ordinary hypergeometric function of
one variable, which is $x=x_1$ now (compare with (\ref{onevar})).
The ordinary hypergeometric function satisfies well-known hypergeometric 
equation
\be\label{ord-eq}
\left(\partial_{x}-{}_pr_s(D)\right)
{}_pF_s(a_1,\dots,a_p;b_1,\dots ,b_s; x)=0,
\quad D:=x\partial_{x} .
\ee
This relation helps us to understand the meaning of function $r$.
\\ 
It is known that the series (\ref{tauzon}) 
diverges if $p>s+1$ (untill any of $a_i+M$ is nonpositive integer).
In case $p=s+1$ it converges in certain domain in the vicinity of 
${\bf x}_{(N)}={\bf 0}$.
For $p<s+1$ the series (\ref{tauzon}) converges for all ${\bf x}_{(N)}$.
These known facts (see \cite{KV}) can be also obtained with the help
of the determinant representation (\ref{det1}) and properties 
of  (\ref{ordinary}).\\

{\bf Example 4}. 
 Hypergeometric function of two
sets of variables ${\bf x}_{(N)},{\bf y}_{(N)}$ we put
\be
 {}_pr_s(D)= 
\frac{\prod_{i=1}^p (a_i+D)}{\prod_{i=1}^s(b_i+D)}
\frac{1}{N-M+D},
\ee
\be
 e^{-T_n}= \frac{1}{\Gamma (N-M+n+1)}
\frac{\prod_{i=1}^p  \Gamma (a_i+n+1)}
{\prod_{i=1}^s \Gamma (b_i+n+1)},
\ee
For the variables ${\bf t}^{+}({\bf x}_{(N)})$ and ${\bf t^*}^{+}({\bf y}_{(N)})$
we obtain (see Section 3 of \cite{Mac} for help) the formula 
(\ref{hZ})
\begin{eqnarray}\label{zon2}
\l M| e^{H({\bf t}^{+}({\bf x}_{(N)}))}e^{A({\bf t^*}^{+}({\bf y}_{(N)}))}|M \r=
{}_p{\mathcal{F}}_s\left.\left(a_1+M,\dots ,a_p+M\atop b_1+M,\dots ,
b_s+M\right|q,{\bf x}_{(N)},{\bf y}_{(N)}\right)=\nonumber\\
\sum_{{\bf n}\atop l({\bf n})\le N }
\frac{s_{{\bf n}}({\bf x}_{(N)})s_{{\bf n}}({\bf y}_{(N)})}
{(N)_{{\bf n}}}
\frac{(a_1+M)_{{\bf n}}\cdots (a_p+M)_{{\bf n}}}
{(b_1+M)_{{\bf n}}\cdots (b_s+M)_{{\bf n}}}. 
 \end{eqnarray}
\\
{\bf Example 5} 
The $q$-generalization of the {\em Example 3}:
\be\label{rq}
{}_pr_s^{(q)}(D)= \frac{\prod_{i=1}^p (1-q^{a_i+D})}
{\prod_{i=1}^s(1-q^{b_i+D})} .
\ee
For the variables ${\bf t}^{+}({\bf x}_{(N)})$ and 
\be\label{chvy}
y_k=q^{k-1},\quad k=1,2,... ,\quad 
t^*_m=\sum_{k=1}^{+\infty} \frac{y_k^m}{m}=
\frac{1}{m(1-q^m)},\quad m=1,2,\dots
\ee
we get Milne's hypergeometric function (\ref{vh}):
\begin{eqnarray}
\l M|e^{H({\bf t})}e^{A({\bf t}^*)}|M \r=
{}_p\Phi_s\left.\left(a_1+M,\dots ,a_p+M\atop b_1+M,\dots ,b_s+M\right|q,
{\bf x}_{(N)}\right)=\nonumber\\
\sum_{{\bf n}\atop l({\bf n})\le N }
\frac{(q^{a_1+M}; q)_{{\bf n}}\cdots (q^{a_p+M};q)_{{\bf n}}}
{(q^{b_1+M};q)_{{\bf n}}\cdots (q^{b_s+M};q)_{{\bf n}}}
\frac{q^{n({\bf n})}}{H_{{\bf n}}(q)}s_{{\bf n}}({\bf x}_{(N)}) .
\label{qtau}
 \end{eqnarray}

{\bf Example 6}.
 To obtain Milne's hypergeometric function of two
sets of variables ${\bf x}_{(N)},{\bf y}_{(N)}$ we use
 ${\bf t}^+({\bf x}_{(N)})$  and ${\bf t^*}^+({\bf y}_{(N)})$.
This choice restricts the sum over partitions  ${\bf n}$ with 
$l{\bf n}\le N$. We put
\be\label{rMilne2}
 {}_pr_s^{(q)}(n)= 
\frac{\prod_{i=1}^p (1-q^{a_i+n})}{\prod_{i=1}^s(1-q^{b_i+n})}
\frac{1}{1-q^{N-M+n}},
\ee
\begin{eqnarray}
 e^{-T_n}= \frac{1}{(1-q)^n\Gamma_q(n+N-M+1)}
\frac{\prod_{i=1}^p (1-q)^n \Gamma_q (a_i+n+1)}
{\prod_{i=1}^s (1-q)^n\Gamma_q(b_i+n+1)},\\
\Gamma_q(a)=(1-q)^{1-a}\frac{(q;q)_{\infty}}{(q^a,q)_{\infty}}, 
\qquad (q^a,q)_n=(1-q)^n\frac{\Gamma_q(a+n)}{\Gamma_q(a)}.
\end{eqnarray}
Here $\Gamma_q(a)$ is a $q$-deformed Gamma-function
\be
\Gamma_q(a)=(1-q)^{1-a}\frac{(q;q)_{\infty}}{(q^a,q)_{\infty}}, 
\qquad (q^a,q)_n=(1-q)^n\frac{\Gamma_q(a+n)}{\Gamma_q(a)}.
\ee
We obtain (see Section 3 of \cite{Mac} for help) the Milne's formula 
(\ref{vh2})
\begin{eqnarray}\label{Milne2}
\tau_r(M,{\bf t}^+({\bf x}_{(N)}), {\bf t^*}^+({\bf y}_{(N)}))=
{}_p\Phi_s\left.\left(a_1+M,\dots ,a_p+M\atop b_1+M,\dots ,b_s+M\right|q,
{\bf x}_{(N)},{\bf y}_{(N)}\right)=\nonumber\\
\sum_{{\bf n}\atop l({\bf n})\le N }
\frac{q^{n({\bf n})}}{H_{{\bf n}}(q)}\frac{s_{{\bf n}}({\bf x}_{(N)})
s_{{\bf n}}({\bf y}_{(N)})}{s_{{\bf n}}(1,q,\dots ,q^{N-1})}
\frac{(q^{a_1+M}; q)_{{\bf n}}\cdots (q^{a_p+M};q)_{{\bf n}}}
{(q^{b_1+M};q)_{{\bf n}}\cdots (q^{b_s+M};q)_{{\bf n}}}. \label{qtauN}
 \end{eqnarray}
This is the KP tau-function (but not the TL one because (\ref{rMilne2})
depends on TL variable $M$).

To receive the basic hypergeometric function of one set of variables 
 we must put indeterminates ${\bf y}_{(N)}$ in (\ref{qtau}) as $y_i=q^{i-1}, i=(1,\dots ,N)$. Thus we have 
\begin{eqnarray}\label{qssN}
{}_p\Phi_s\left.\left(a_1+M,\dots ,a_p+M\atop b_1+M,\dots ,b_s+M\right|q,{\bf x}_{(N)}\right)
=\nonumber\\ \sum_{{\bf n}\atop l({\bf n})\le N }\frac {(q^{a_1+M};q)_{{\bf n}}\cdots (q^{a_p+M};q)_{{\bf n}}}
                            {(q^{b_1+M};q)_{{\bf n}}\cdots (q^{b_s+M};q)_{{\bf n}}}
\frac{q^{n({\bf n})}}{H_{{\bf n}}(q)}s_{{\bf n}}({\bf x}_{(N)}) .
\end{eqnarray}
And for $N=1$ we have the ordinary $q$-deformed hypergeometrical function:
\begin{eqnarray}\label{qssN=1}
{}_p\Phi_s\left.\left(a_1+M,\dots ,a_p+M\atop b_1+M,\dots ,b_s+M\right|q,x\right)
=\sum_{n=0}^{+\infty}\frac{(q^{a_1+M};q)_{n}
\cdots (q^{a_p+M};q)_{n}}{(q^{b_1+M};q)_{n}\cdots (q^{b_s+M};q)_{n}}
\frac{x^{n}}{(q;q)_n}, \quad x=x_1 \quad
\end{eqnarray}
which satisfies the  $q$-difference equation (compare it with
(\ref{ord-eq})
\be\label{q-eq}
\left(\frac {1}{x} \left(1-q^{D}\right)-{}_p r_s^{(q)}(D)\right)
{}_p\Phi_s(a_1,\dots ,a_p;b_1,\dots ,b_s;q,x)=0,
\quad D:=x\partial_{x} ,
\ee
where ${}_p r_s^{(q)}(D)$ is defined by (\ref{rq}).

For the bosonic representation of hypergeometric function (\ref{qssN=1}) 
see \cite{MV}.\\   
There are various applications for series (\ref{qssN=1}), for instance see
\cite{QCD},\cite{Pugai} and \cite{Odz}. Bosonic representation of 
(\ref{qssN=1}) was found in \cite{MV}.
 Let us note that operator $q^D$ which acts on fermions $\psi(z)$
was used in \cite{LS} in different context.\\

{\bf Example 7} 

Notations and notions for this {\em Example} we borrowed from \cite{Zab}.
Let $r$ be a rational function of Jackoby theta-functions 
$\theta (2x\eta |{\tau}')$, where $\tau '$ is an elliptic 
modulus:
\begin{eqnarray}
{}_pr_s^{(\eta)}(n)= 
\frac{\prod_{i=1}^p \theta(2\eta({a_i+n})|\tau ')}
{\theta(2\eta({N-M+n})|\tau ')
\prod_{i=1}^s
\theta(2\eta({b_i+n})|\tau ')},
e^{-T_{n-1}}=\frac {\prod_{i=1}^p [a_i]_n}
{[N-M]_n\prod_{i=1}^s [b_i]_n}.
\end{eqnarray}
Here elliptic Pochhammer's symbol $[a]_n$  is
defined in terms of the elliptic number [a]
\be
[a]=\theta (2a\eta |\tau '),\quad [a]_k=[a][a+1][a+2]\cdots [a+k-1] .
\ee
One can associate the elliptic Pochhammer's symbol with a 
given partition ${\bf n}$: 
\be
[a]_{\bf n}=[a]_{n_1}[a-1]_{n_2}\cdots [a-l+1]_{n_l} .
\ee
For the variables ${\bf t}^+({\bf x}_{(N)})$ and ${\bf t^*}^+({\bf y}_{(N)})$
we can introduce the hypergeometric function
\begin{eqnarray}
\l M| e^{H({\bf t}^+({\bf x}_{(N)}))}e^{A({\bf t^*}^+({\bf y}_{(N)}))}|M \r=
{}_pF_s^{(\eta)}\left.\left(a_1+M,\dots ,a_p+M\atop b_1+M,\dots ,
b_s+M\right|\eta,{\bf x}_{(N)},{\bf y}_{(N)}\right)=\nonumber\\
\sum_{{\bf n}\atop l({\bf n})\le N }
\frac{s_{{\bf n}}({\bf x}_{(N)})s_{{\bf n}}({\bf y}_{(N)})}
{[N]_{{\bf n}}}
\frac{[a_1+M]_{{\bf n}}\cdots [a_p+M]_{{\bf n}}}
{[b_1+M]_{{\bf n}}\cdots [b_s+M]_{{\bf n}}}. \label{qtauNe}
 \end{eqnarray}
As in the case of (\ref{vh}) this the KP tau-function which is
not the TL tau-function because the factor $[N]_{{\bf n}}$ in the 
denominator does not depend on $M$.
For $N=1$ we get elliptic hypergeometric function
of one variable \cite{Zab}.
For instance to obtain the elliptic very-well-poised hypergeometric function
\be
{}_{p+1}W_p(\alpha_1;\alpha_4,\alpha_5,\dots  ,\alpha_{p+1};z|\eta,\tau ')
=\sum_{n=0}^\infty z^n \frac{[\alpha_1+2n][\alpha_1]_n}{[\alpha_1][n]!}
\prod_{m=1}^{p-2}\frac{[\alpha_{m+3}]_n}{[\alpha_1-\alpha_{m+3}+1]_n} ,
\ee
we choose
\be
t_n=\frac{z^n}{n} ,\quad t^*_n=\frac{1}{n} ,\quad
e^{-T_{n-1}}=\frac{[\alpha_1+2n][\alpha_1]_n}{[\alpha_1][n]!}
\prod_{m=1}^{p-2}\frac{[\alpha_{m+3}]_n}
{[\alpha_1-\alpha_{m+3}+1]_n} .
\ee

{\bf Example 8} The hypergeometric function (\ref{vh2}),(\ref{Milne2}) 
${}_1\Phi_1\left.\left(a \atop b \right|q,
{\bf x}_{(N)},{\bf y}_{(N)}\right)$
 can be degenerated to
${}_1\Phi_0\left.\left(a  \right|q,
{\bf x}_{(N)},{\bf y}_{(N)}\right)$ by taking $b \to +\infty$ (remember that 
$|q|<1$).
The limit $b \to -\infty$ (with the rescaling of times $x_i,y_i \to
q^{\frac b2}x_i,q^{\frac b2}x_i$) is also of interest. Consider this 
limit and put $a=N-M$.

Now we get an example of KP tau function 
(\ref{tauhyp}) which is not a hypergeometric function. 
Take $T_n=-\frac {\g}{2} (n+\frac 12 )^2$,  or the same
\be\label{rokunkov}
r(D)=q^{-D},\quad q=e^{-\g},
\ee
and rescale the times once more:
$t_k={\alpha}^kp_k,t_k^*={\alpha}^kp_k^*$.
We get the series
\begin{eqnarray}\label{okunkov}
\l M |e^{H({\bf t})}e^{A({\bf t}^*)} |M \r
= \sum_{\bf n} {\alpha}^{|{\bf n}|}
e^{\g f_2({\bf n })}s_{\bf n }({\bf p})s_{\bf n }({\bf p}^* ) ,\\
 f_2({\bf n })=\frac12 \sum_i \left[(n_i-i+M-\frac12)^2-(-i+M-\frac12)^2
\right],
\end{eqnarray}
 which was recently considered in \cite{O2} (our notations 
${\bf n},\alpha ,\g $
are related to $\lambda ,q,\beta $ in \cite{O2} respectively). This
series is a generating function for double Hurwitz numbers $Hur_{d,b}(n,m)$
introduced in \cite{O2} as follows. $Hur_{d,b}(n,m)$ is a weighted 
number of connected 
degree $d$ covering of $P^1$ with monodromy around $0,\infty \in P^1$
being $n$ and $m$,  respectively, and $b$ additional simple ramifications.
The genus of each covering is $g=(b+2-l(n)-l(m))/2$, where $l(n)$ is the 
number of parts of $n$.  The weight of each covering is the reciprocal of 
the order of its automorphism group.
The formula presented in \cite{O2} in our terms reads as
\be
\log \l M |e^{H({\bf t})}e^{A({\bf t}^*)} |M \r =
\sum_{d,b,n,m}\alpha^d \g^b p_n p_m^* Hur_{d,b}(n,m)/b! ,
\ee
for $A$ see (\ref{Abeta}),(\ref{dopsim}),(\ref{rokunkov}). Therefore
the generating function for the double Hurwitz numbers is expressed
in terms of group cocycle of the $\Psi DO$ on the circle (see
{\em Appendix ``Gauss factorization problem, additional symmetries, 
string equations and $\Psi DO$ on the circle}), which is
the correlator under the logarithm.
\\
{\bf Example 9}. Take $r$ be a step function: $r(n)=0,n<k$ and
$r(n)=1,n \geq k$, and let $k<N$. Then
\be\label{Gessel}
\tau_r({\bf x}_{(N)},{\bf y}_{(N)})=\sum_{{\bf n}\atop l({\bf n})\leq k<N}s_{\bf n}({\bf x}_{(N)})
s_{\bf n}({\bf y}_{(N)})
\ee
is equal the determinant of a Toeplitz matrix - it is a subject of Gessel's 
theorem. We have the determinant representation of (\ref{Gessel}) due to 
the formula (\ref{det1}).

\subsection{Baker-Akhiezer functions and Sato Grassmannian}
Let us write down the expression for Baker-Akhiezer functions
(\ref{baker}) in terms of Miwa variables (\ref{Miwa-}):
\be\label{bak}
w_{\infty}(M,{\bf t}^{-}({\bf x}_{(N)}),{\bf t^*},\frac{1}{z})=
\frac{\tau_r(M,{\bf t}^{-}({\bf x}_{(N+1)}),{\bf t^*})}{\tau_r(M,{\bf t}^{-}({\bf x}_{(N)}),{\bf t^*})}
\prod_{i=1}^N(1-\frac{x_i}{z}),\quad
{\bf x}_{(N+1)}=(x_1,\dots ,x_N,z) ,
\ee
\be\label{bak*}
w_{\infty}^*(M,{\bf t},{\bf t^*},\frac{1}{z})=
-\frac{\tau_r(M,{\bf t+[z]},{\bf t^*})}{\tau_r(M,{\bf x}_{(N)},{\bf t^*})}
\prod_{i=1}^N\frac{1}{(1-\frac{x_i}{z})}\frac{dz}{z},\quad
{\bf [z]}=(z,\frac{z^2}{2},\dots).
\ee

We see that the variables $x_k,k=1,\dots,N$ are zeroes of 
$w_{\infty}(z)$ and poles of
$w_{\infty}^*(z)$.\\
\bremark
The associated linear problems for Baker-Akhiezer functions are read as
\be\label{Loperator}
\left(\partial_{t_1}-\partial_{t_1}\phi_n\right)
w(n,{\bf t},{\bf t}^*,z)
=w(n+1,{\bf t},{\bf t}^*,z),
\ee
\be\label{Aoperator}
\partial_{t_1^*}
w(n,{\bf t},{\bf t}^*,z)=r(n)e^{\phi_{n-1}-\phi_{n}}
w(n-1,{\bf t},{\bf t}^*,z).
\ee
where $w$ is either $w_\infty $ or $w_0$.
The compatibility of these equations gives rise to the equation 
(\ref{rToda}). Taking into account the second eq.(\ref{rToda}), 
equations (\ref{Loperator}), (\ref{Aoperator})
may be also viewed as the recurrent equations for the tau-functions
which depend on different number of variables ${\bf x}_{(N)}$.
\eremark
Let us write down a plane of Baker-Akhiezer functions (\ref{baker}), which 
characterizes Sato Grassmannian related to the tau-function (\ref{tau3})
$\tau_r(M,{\bf t},{\bf t}^*)$.
We take $x_k=0,k=1,\dots,N$ in (\ref{bak}) and obtain:
\be\label{SatoG}
w_\infty (n,{\bf 0},{\bf t}^*,z)=z^n(1+\sum_{m=1}^\infty 
r(n)r(n-1)\cdots r(n-m+1)p_m(-{\bf t}^*)z^{-m})
, \quad n=M,M+1,M+2,... .
\ee
The dual plane is
\be\label{dSatoG}
w_\infty^* (n,{\bf 0},{\bf t}^*,z)=z^{-n}(1+\sum_{m=1}^\infty 
r(n)r(n+1)\cdots r(n+m-1) p_m({\bf t}^*)z^{-m})dz , \quad n=M,M+1,M+2,... .
\ee
About these formulae see also (\ref{SatoGr}),(\ref{dSatoGr}).\\
We see that when $r$ has zeroes, then in the regions (\ref{finitechainh})
the Grassmannian is the finite-dimensional one. The corresponding 
tau-function is a particular case of the one found in \cite{LSav},
\cite{Nak}.

\subsection{Different representations}

Let us rewrite  hypergeometric series in different way representing 
all Pochhammer's coefficients $(q^a;q)_{\bf n}$ and $(a)_{\bf n}$  
 through Schur functions.  This gives us the opportunity to
interchange the role of Pochhammer's coefficients and Schur
functions in (\ref{vh2}),(\ref{beskshur}), and to present different 
fermionic representations of the hypergeometric functions.
We have the  relations (see \cite{Mac}):
\be\label{PochSchur}
\prod_{(i,j)\in {\bf n}} (1-q^{a+j-i})=
\frac{s_{{\bf n}}({\bf t}(a,q))}{s_{{\bf n}}({\bf t}(+\infty,q))},\quad 
\prod_{(i,j)\in {\bf n}}(a+j-i)=\frac{s_{{\bf n}}({\bf t}(a))}{s_{{\bf n}}({\bf t}(+\infty)} ,
\ee
where parameters $t_m(a,q)$ 
and $t_m(a)$ are chosen via generalized Miwa transform \cite{Miwa} 
with multiplicity $a$  (remember that $|q|<1$)
\begin{eqnarray}\label{t(a)}
t_m(a,q)=\frac{1-(q^a)^{m}}{m(1-q^m)},\quad  
t_m(a)=\frac{a}{m},
\quad m=1, 2,\dots ,\\
 s_{{\bf n}}({\bf t}(+\infty,q))=\lim_{a\to +\infty}
s_{{\bf n}}({\bf t}(a,q))=\frac{q^{n({\bf n})}}{H_{{\bf n}}(q)} ,\\
 s_{{\bf n}}({\bf t}(+\infty))=\lim_{a\to +\infty}
s_{{\bf n}}\left(\frac{t_1(a)}{a}, \frac{t_2(a)}{a^2},\dots\right)=
\lim_{a\to +\infty}\frac{1}{a^{|{\bf n}|}}s_{{\bf n}}({\bf t}(a))=
\frac{1}{H_{{\bf n}}} .
\end{eqnarray}
Now we rewrite the series (\ref{qtau}) and (\ref{beskshur}) only in terms
of Schur functions:
\begin{eqnarray}\label{qtaushur}
{}_p\Phi_s\left.\left(a_1+M,\dots ,a_p+M\atop b_1+M,
\dots ,b_s+M\right|q,{\bf x}_{(N)},{\bf y}_{(N)}\right)=
\tau_{r}(M,{\bf t}(+\infty,q),{\bf t}^*) \nonumber\\ 
=\sum_{{\bf n}\atop l({\bf n})\le N }
\frac{\prod_{k=1}^p{s_{{\bf n}}({\bf t}(a_k+M,q))}}
{\prod_{k=1}^s{s_{{\bf n}}({\bf t}(b_k+M,q))}}
\left(s_{{\bf n}}({\bf t}(+\infty,q))\right)^{s-p+1}
\frac
{s_{{\bf n}}({\bf x}_{(N)})s_{{\bf n}}({\bf y}_{(N)})}
{s_{{\bf n}}({\bf t}(N,q))},
\end{eqnarray}
\begin{eqnarray}\label{taushur}
{}_p{\mathcal{F}}_s\left.\left(a_1+M,\dots ,a_p+M\atop b_1+M,
\cdots ,b_s+M\right| {\bf x}_{(N)},{\bf y}_{(N)}\right)=
\tau_{r}(M,{\bf t}(+\infty),{\bf t})=\nonumber\\
\sum_{{\bf n}\atop l({\bf n})\le N }
\frac{\prod_{k=1}^p s_{{\bf n}}({\bf t}(a_k+M))}
{\prod_{k=1}^s s_{{\bf n}}({\bf t}(b_k+M))}
\left(s_{{\bf n}}({\bf t}(+\infty))\right)^{s-p+1}\frac
{s_{{\bf n}}({\bf x}_{(N)})s_{{\bf n}}({\bf y}_{(N)})}
{s_{{\bf n}}({\bf t}(N))}.
\end{eqnarray}
A nice feature of this formulae is that they do not contain
number coefficients at all, it is a sum of ratios of Schur
functions only.

We obtain different fermionic representations of hypergeometric 
functions (\ref{taushur}), (\ref{qtaushur}), and they are parametrized
by a complex noninteger number $b$:\\
\bprop
For $b \in C$ and for $r= {}_pr_s$ 
(see (\ref{op1})) we have
\be\label{sym=miwa}
\tau_r(M,{\bf t}(+\infty),{\bf t}^*)=\tau_{r_b}(M,{\bf t}(b+M),{\bf t}^*),
\quad r_b=\frac{r}{b+D}.
\ee
For $r= {}_pr_s^{(q)}$ (see (\ref{rq})) we have
\be\label{qsym=miwa}
\tau_r(M,{\bf t}(+\infty,q),{\bf t}^*)=\tau_{r_b}(M,{\bf t}(b+M,q),{\bf t}^*),
\quad r_b=\frac{r}{1-q^{b+D}} .
\ee
\eprop
\bremark
 There are two ways to restrict the sum 
(\ref{tauhyp}) to the sum over partitions of length $l({\bf n})\leq N$.
 First, if we use Miwa's change (\ref{Miwa+}), then
$s_{{\bf n}}({\bf x}_{(N)})=0$, for ${\bf n}$ with length $l({\bf n})>N$.
The second way is to restrict the  Pochhammer's coefficients:
if we put  $a_i=N$ for one $i$ from (\ref{rq}) equal to $N$, then the 
coefficient $(q^{a_i},q)_{{\bf n}}$ vanishes for  $l({\bf n})>N$.
Since we expressed Pochhammer's coefficients in terms of Schur functions
in (\ref{PochSchur}) both ways have the same explanation.
Indeed
\be
t_m(N,q)=\frac{1}{m}\frac{1-(q^N)^m}{1-q^m}=
\frac{1}{m}(1+(q)^m+(q^2)^m+\cdots+(q^{N-1})^m) .
\ee
Therefore we obtain for Miwa's change: $x_1=1,x_2=q,\dots , x_N=q^{N-1}$ and
\be
s_{{\bf n}}({\bf t}(N,q))=s_{{\bf n}}(1, q,\dots, q^{N-1})=0,
\quad l({\bf n})>N.
\ee
The same we have for the sum
over partitions ${\bf n}$ such that $l({\bf n'})<K$.
Again the first way has to be realized through the following Miwa's 
change of variables:
\be\label{mtmiwa}
t_m=-\sum_{i=1}^{K}\frac{x_i^m}{m},\quad s_{{\bf n}}({\bf t})=s_{{\bf n'}}({\bf x}_{(K)}) .
\ee
The second way is to make one of the parameters, for example  $a_j$ 
from (\ref{rq}) equal to $(-K)$. In this case
\be
s_{{\bf n}}({\bf t}(-K,q))=s_{{\bf n'}}\left(\frac{1}{q},\frac{1}{q^2},\dots,\frac{1}{q^{K}}\right)=0, \quad l({\bf n'})>K.
\ee
\eremark

\section{ Further generalization. Examples of Gelfand-Graev hypergeometric
functions}
\subsection{Generalization}
Formula (\ref{teplitsGauss}) is related to 'Gauss
decomposition'
of operators inside vacuums $\langle 0|\dots|0\rangle $ into
diagonal operator $e^{H_0({\bf T})}$ and
upper triangular operator $e^{H({\bf t})}$  and lower triangular
operator $e^{-H^*({\bf t}^*)}$ the last two have the Toeplitz form.
Now let us consider more general two-dimensional 
Toda chain tau-function
\begin{equation}\label{general}
\tau=\langle M|e^{H({\bf t})}ge^{-A({\bf t}^*)}|M\rangle ,
\end{equation}
where  we decompose $g$ in the following way:
\begin{equation}\label{general1}
g(\tilde{\gamma},\gamma)=e^{\tilde{A}_1(\tilde{\gamma}_1)}
\cdots e^{\tilde{A}_k(\tilde{\gamma}_k)}e^{-A_l(\gamma_l)}
\cdots e^{-A_1(\gamma_1)},
\end{equation}
where each of $\tilde{\gamma_i},\gamma_i,\tilde{A}_i,A_i$ has an additional 
index:  $\tilde{\gamma}_{in},\gamma_{in},\tilde{A}_{in},A_{in}$,
($n=1,2,\dots$).
Here each of $A_i(\gamma_i)$ has a form as in (\ref{dopsim}),(\ref{Abeta}) and corresponds to operator $r^i(D)$, while each of $\tilde{A}_i(\tilde{\gamma }_i)$ has a form of  (\ref{tdopsim})  and corresponds to operator $\tilde{r}^i(D)$.
 Collections of variables
$\tilde{\gamma}=\{\tilde{\gamma}_{in}\},
\gamma=\{\gamma_{in}\}$ play the role of coordinates for some wide enough 
class of Clifford group elements $g$.
This tau-function is related to rather involved
generalization of the hypergeometric functions we considered above.
Tau-function (\ref{general}),(\ref{general1}) may be considered
as the result of applying of the additional symmetries to the vacuum
tau function, which is 1, see {\em Appendix ``The vertex operator action''}.\\
Let us calculate this tau-function. First of all we introduce a set consisting of $m+1$ partitions:
\be
({\bf n_1},\dots ,{\bf n_m}, {\bf n_{m+1}}={\bf n}),\quad 
0\le {\bf n_1}\le {\bf n_2}\le \cdots\le {\bf n_m}\le {\bf n_{m+1}}={\bf n},
\ee
see \cite{Mac} for the notation $\le$ for the partitions. 
The corresponding set 
\be
\Theta_{{\bf n}}^m=({\bf n_1},\theta_1,\dots ,\theta_{m}),\quad \theta_i={\bf n_{i+1}}-{\bf n_{i}},\quad i=1,\dots ,m
\ee
depends on the partition ${\bf n}$ and the number  $m+1$ of the 
partitions.  We take 
as $s_{\Theta}({\bf t}^*,\gamma)$  the product which is relevant to the set 
$\Theta^m_{{\bf n}}$ and depending on the set of variables 
$\mu_{i}=\{\mu_{ij}\}$ ($i=(1,\dots , m+1)$, $j=(1,2\dots)$)  
\begin{eqnarray}
s_{\Theta^m_{{\bf n}}}({\bf \mu})=s_{{\bf n_1}}(\mu_1)s_{\theta_1}(\mu_2)\cdots s_{\theta_m}(\mu_{m+1}).
\end{eqnarray}
Here $s_{\theta_i}$ is a skew Schur function (see \cite{Mac}). Further we define function $r_{\Theta^m_{{\bf n}}}(M)$:
\be
r_{\Theta^m_{{\bf n}}}(M)=r_{{\bf n_1}}(M)r^{1}_{\theta_1}(M)\cdots r^{m}_{\theta_m}(M),
\ee
where the function $r^i_{\theta_i}(M)$ , a skew analogy of $r_{\bf n}(M)$ from (\ref{r_n}), is
\be\label{rskew}
r_{\theta_i}(M)=\prod_{j=1}^s r(n^{(i)}_{j}-j+1+M)\cdots r(n^{(i+1)}_{j}-j+M),
\ee 
where ${\bf n_{i+1}}=(n^{(i+1)}_{1},\dots ,n^{(i+1)}_{s})$. If the function $r^i(m)$  has no poles and zeroes at integer points then the relation
\be
r^{i}_{\theta_i}(M)=\frac{r^{i}_{{\bf n_{i+1}}}(M)}{r^{i}_{{\bf n_{i}}}(M)},\quad i=1,\dots , m
\ee 
is correct.
To calculate the tau function we need the 
{\em Lemma}\\
{\bf Lemma 3} Let partitions ${\bf n}=(i_1,\dots ,i_s|j_1-1,\dots ,j_s-1)$ and
${\bf \tilde{n}}=(\tilde{i}_1,\dots ,\tilde{i}_r|\tilde{j}_1-1,\dots ,\tilde{j}_r-1)$ satisfy the relation ${\bf n}\ge{\bf \tilde{n}}$.
The following  is valid:
\begin{eqnarray}
\l 0|\psi^*_{\tilde{i}_1}\cdots\psi^*_{\tilde{i}_r}\psi_{-\tilde{j}_r}\cdots
\psi_{-\tilde{j}_1}e^{A^i(\g_i)}\psi^*_{-j_1}\cdots
\psi^*_{-j_s}\psi_{i_s}\cdots\psi_{i_1}|0\r=\nonumber\\
=(-1)^{\tilde{j}_1+\cdots+\tilde{j}_r+j_1+\cdots+j_s}s_{\theta}(\g_i)r_{\theta}(0),\qquad \theta={\bf n}-\tilde{{\bf n}}.
\end{eqnarray}

{\bf Proof:} the proof is achieved by direct calculation (see Example 22 in
Sec 5 of \cite{Mac} for help).\\
Then we obtain the  generalization of 
{\em Proposition 1}:\\
\bprop
\be\label{gen}
\tau_M({\bf t},{\bf t}^* ;\g , \tilde{\g})=\sum_{{\bf n}}
\sum_{\Theta^k_{{\bf n}}}
\sum_{\Theta^l_{{\bf n}}} \tilde{r}_{\Theta^k_{{\bf n}}}(M)
r_{\Theta^l_{{\bf n}}}(M)
s_{\Theta^k_{{\bf n}}}({\bf t},\tilde{\g}) s_{\Theta^l_{{\bf n}}}({\bf t}^*,\g) ,
\ee 
where $\tilde{r}_{\Theta^k_{{\bf n}}}(M)$ and $r_{\Theta^l_{{\bf n}}}(M)$ are
given by (\ref{rskew}).
\eprop

With the help of this series one can obtain different hypergeometric
functions.

\subsection{The example of Gelfand,Graev and Retakh hypergeometric
series}
Let us consider the  tau function:
\be
\tau(M,\tilde{\beta},\beta;\g)=
\l M|e^{\tilde{A}(\tilde{\beta})}e^{-A_l(\gamma_l)}
\cdots e^{-A_1(\gamma_1)}e^{-A(\beta)}|M\r .
\ee
We put 
\be
 \tilde{\beta}=(x, \frac{x^2}{2},\frac{x^3}{3},\dots),\quad \beta=(y_1,0,0,\dots ),\quad \g_i =(y_{i+1},0,0,\dots )\quad i=(1,\dots,l).
\ee
We obtain the series
\begin{eqnarray}\label{Horn}
\tau(M, x, y_1,\dots,y_{l+1})=\sum_{n_1,\dots, n_{l+1}=0}^{+\infty}
\tilde{r}_{(n_1+\cdots+n_{l+1})}(M) r_{\Theta^l_{{\bf n}}}(M)\frac{(xy_1)^{n_1}\cdots (xy_{l+1})^{n_{l+1}}}{n_1!\cdots n_{l+1}!}=\qquad \qquad \\
\sum_{n_1,\dots,n_{l+1}\in Z}c(n_1,\dots,n_{l+1}) (xy_1)^{n_1}\cdots (xy_{l+1})^{n_{l+1}}, c(n_1,\dots,n_{l+1})=\frac{\tilde{r}_{(n_1+\cdots+n_{l+1})}(M) r_{\Theta^l_{{\bf n}}}(M)}{\Gamma(n_1+1)\cdots\Gamma(n_{l+1}+1)},\quad
\end{eqnarray}
where $\Theta^l_{{\bf n}}$ corresponds to the set of simple partitions-rows 
\be
{\bf n_1}=(n_1),{\bf n_2}=(n_1+n_2),\dots, {\bf n}_{l+1}=(n_1+\cdots+n_{l+1})
\ee
When functions $b_i(n_1,\dots,n_{l+1})$ defined as
\be
b_i(n_1,\dots,n_{l+1})=\frac{c(n_1,\dots, n_i+1,\dots, n_{l+1})}{c(n_1,\dots, n_{l+1})},\quad i=1,\dots, l+1
\ee
are rational functions of $(n_1,\dots, n_{l+1})$, then tau function 
(\ref{Horn}) is a Horn hypergeometric series \cite{KV}. \\
Above series for the special choice of functions $r^i(D)$ can be deduced from the
Gelfand, Graev and Retakh series defined on the special lattice and corresponding to the special set of parameters. Let us take the  rational functions $r^i(D)$:
\begin{eqnarray}
r^i(D)=\frac{\prod_{j=1}^{p^{(i)}}(D+a^{(i)}_j)}{\prod_{m=1}^{s^{(i)}}(D+b^{(i)}_m)},\quad (i=0,\dots , l),\qquad
r^0(D)=r(D)\\
\tilde{r}(D)=\frac{\prod_{j=1}^{p^{(l+1)}}(D+a^{(l+1)}_j)}{\prod_{m=1}^{s^{(l+1)}}(D+b^{(l+1)}_m)}
\end{eqnarray}     
Let define $N=p^{(0)}+s^{(0)}+2\sum_{j=1}^l (p^{(j)}+2s^{(j)})+p^{(l+1)}+s^{(l+1)}+l+1$
and consider complex space $C^N$. In this space we consider the $l+1$-dimensional basis $B$ and the vector $\upsilon$ consisting of parameters. 
\begin{eqnarray}
p_0=s_0=0,\quad p_i=p^{(i-1)}+s^{(i)},\quad s_i=s^{(i-1)}+p^{(i)},\quad i=(1,\dots , l)\nonumber\\
p_{l+1}=p^{(l)}+p^{(l+1)},\quad s_{l+1}=s^{(l)}+s^{(l+1)},\quad  N=\sum_{j=1}^{l+1} (p_j+s_j)+l+1 
\end{eqnarray}
\begin{eqnarray}\label{f_i}
{\bf f}^i=-({\bf e}_{p_0+s_0+\cdots +p_{i-1}+s_{i-1}+1}+\cdots +{\bf e}_{p_1+s_1+\cdots +p_{i-1}+s_{i-1}+p_i})+\nonumber\\
+({\bf e}_{p_1+s_1+\cdots +p_{i-1}+s_{i-1}+p_i+1}+\cdots +{\bf e}_{p_1+s_1+\cdots +p_{i-1}+s_{i-1}+p_i+s_i}), \quad i=1,\dots ,l+1
\end{eqnarray}
where ${\bf e}_{i}=\underbrace{(0,\dots , 0,\stackrel{i}{\hat{1}}, 0,\dots)}_{\mbox{N}}$. 
The lattice $B\in C^N$ is generated by the vector basis of dimension $l+1$:
\be
{\bf b}^i={\bf f}^i+\cdots +{\bf f}^{l+1}+{\bf e}_{N-l-1+i},\qquad i=1,\dots ,l+1
\ee
Vector $\upsilon\in C^N$ is defined as follows (compare with (\ref{f_i})):
\begin{eqnarray}
\upsilon^i =-(a_1^{(i-1)}{\bf e}_{p_0+s_0+\cdots +p_{i-1}+s_{i-1}+1}+\cdots +a^{(i-1)}_{p^{(i-1)}}{\bf e}_{p_0+s_0+\cdots +p_{i-1}+s_{i-1}+p^{(i-1)}}+\nonumber\\
+b_1^{(i)}{\bf e}_{p_0+s_0+\cdots +p_{i-1}+s_{i-1}+p^{(i-1)}+1}+\cdots +b^{(i)}_{s^{(i)}}{\bf e}_{p_0+s_0+\cdots +p_{i-1}+s_{i-1}+p_i})+\nonumber\\
+((b^{(i-1)}_1-1){\bf e}_{p_0+s_0+\cdots +p_{i-1}+s_{i-1}+p_i+1}+\cdots +
(b^{(i-1)}_{s^{(i-1)}}-1){\bf e}_{p_0+s_0+\cdots +p_{i-1}+s_{i-1}+p_i+s^{(i-1)}}+
\nonumber\\
+(a^{(i)}_1-1){\bf e}_{p_0+s_0+\cdots +p_{i-1}+s_{i-1}+p_i+s^{(i-1)}+1}+\cdots +
(a^{(i)}_{s^{(i)}}-1){\bf e}_{p_0+s_0+\cdots +p_{i-1}+s_{i-1}+p_i+s_i})
\end{eqnarray}
for $i=(1,\dots l)$, and
\begin{eqnarray}
\upsilon^{l+1} =-(a_1^{(l)}{\bf e}_{p_0+s_0+\cdots +p_{l}+s_{l}+1}+\cdots +a^{(l)}_{p^{(l)}}{\bf e}_{p_0+s_0+\cdots +p_{l}+s_{l}+p^{(l)}}+\nonumber\\
+a_1^{(l+1)}{\bf e}_{p_0+s_0+\cdots +p_{l}+s_{l}+p^{(l)}+1}+\cdots +a^{(l+1)}_{s^{(l+1)}}{\bf e}_{p_0+s_0+\cdots +p_{l}+s_{l}+p_{l+1}})+\nonumber\\
+((b^{(l)}_1-1){\bf e}_{p_0+s_0+\cdots +p_{l}+s_{l}+p_{l+1}+1}+\cdots +
(b^{(l)}_{s^{(l)}}-1){\bf e}_{p_0+s_0+\cdots +p_{l}+s_{l}+p_{l+1}+s^{(l)}}+
\nonumber\\
+(b^{(l+1)}_1-1){\bf e}_{p_0+s_0+\cdots +p_{l}+s_{l}+p_{l+1}+s^{(l)}+1}+\cdots +
(b^{(l+1)}_{s^{(l+1)}}-1){\bf e}_{p_0+s_0+\cdots +p_{l}+s_{l}+p_{l+1}+s_{l+1}})
\end{eqnarray}
Vector $\upsilon$ is:
\be
\upsilon=\upsilon^1+\cdots +\upsilon^{l+1}
\ee
Now we can write down Gelfand, Graev and Retakh hypergeometric series corresponding to the lattice $B$ and vector $\upsilon$:
\begin{eqnarray}
F_B(\upsilon;z)=\sum_{{\bf b}\in B}\prod_{j=1}^N \frac{z_j^{\upsilon_j+b_j}}{\Gamma(\upsilon_j+b_j+1)}+\nonumber\\
\sum_{n_1,\dots , n_{l+1}\in Z}\prod_{j=1}^N
\frac{z_j^{\upsilon_j+n_1b^1_j+\cdots +n_{l+1}b^{l+1}_j}}
{\Gamma(\upsilon_j+n_1b^1_j+\cdots +n_{l+1}b^{l+1}_j+1)}
\end{eqnarray}
Let us compare this series with tau function (\ref{Horn}):
\be
F_B(\upsilon;{\bf z})=c_1(a,b)g_1({\bf z})\cdots c_{l+1}(a,b)g_{l+1}({\bf z})\tau(M, x, y_1,\dots,y_{l+1})
\ee
where
\begin{eqnarray}
c_i^{-1}(a,b) =\Gamma(1-a_1^{(i-1)})\cdots\Gamma(1-a^{(i-1)}_{p^{(i-1)}})
\Gamma(1-b_1^{(i)})\cdots\Gamma(1-b^{(i)}_{s^{(i)}})\times\nonumber\\
\times\Gamma(b^{(i-1)}_1)\cdots\Gamma(b^{(i-1)}_{s^{(i-1)}})
\Gamma(a^{(i)}_1)\cdots\Gamma(a^{(i)}_{s^{(i)}}),\quad i=1,\dots ,l
\end{eqnarray}

\begin{eqnarray}
c_{l+1}^{-1}(a,b) =\Gamma(1-a_1^{(l)})\cdots\Gamma(1-a^{(l)}_{p^{(l)}})\Gamma(1-a_1^{(l+1)})
\cdots\Gamma(1-a^{(l+1)}_{s^{(l+1)}})\times\nonumber\\
\times\Gamma(b^{(l)}_1)\cdots\Gamma(b^{(l)}_{s^{(l)}})\Gamma(b^{(l+1)}_1)
\cdots\Gamma(b^{(l+1)}_{s^{(l+1)}})
\end{eqnarray}
\begin{eqnarray}
\frac{z_{p_0+s_0+\cdots +p_{i-1}+s_{i-1}+p_i+1}\cdots z_{p_1+s_1+\cdots +p_{i-1}+s_{i-1}+p_i+s_i}}{(-z_{p_0+s_0+\cdots +p_{i-1}+s_{i-1}+1})\cdots (-z_{p_0+s_0+\cdots +p_{i-1}+s_{i-1}+p_i})}=1, \quad i=2,\dots ,l
\end{eqnarray}
\begin{eqnarray}
\frac{z_{p_1+1}\cdots z_{p_1+s_1}z_{N-l}}{(-z_{1})\cdots (-z_{p_1})}=y_1
\end{eqnarray}
\be
y_i=z_{N-l-1+i},\quad i=2,\dots ,l+1
\ee
\begin{eqnarray}
\frac{z_{p_1+s_1+\cdots +p_{l}+s_{l}+p_{l+1}+1}\cdots z_{p_1+s_1+\cdots +p_{l}+s_{l}+p_{l+1}+s_{l+1}}}{(-z_{p_1+s_1+\cdots +p_{l}+s_{l}+1})\cdots (-z_{p_1+s_1+\cdots +p_{l}+s_{l}+p_{l+1}})}=x
\end{eqnarray}

\begin{eqnarray}
g_i({\bf z})={z}^{\left(-a_1^{(i-1)}\right)}_{p_0+s_0+\cdots +p_{i-1}+s_{i-1}+1}\cdots {z}^{\left(-a^{(i-1)}_{p^{(i-1)}}\right)}_{p_0+s_0+\cdots +p_{i-1}+s_{i-1}+p^{(i-1)}}\times\nonumber\\
\times {z}^{\left(-b_1^{(i)}\right)}_{p_0+s_0+\cdots +p_{i-1}+s_{i-1}+p^{(i-1)}+1}\cdots{z}^{\left(-b^{(i)}_{s^{(i)}}\right)}_{p_0+s_0+\cdots +p_{i-1}+s_{i-1}+p_i}\times\nonumber\\ 
\times {z}^{\left(b^{(i-1)}_1-1\right)}_{p_0+s_0+\cdots +p_{i-1}+s_{i-1}+p_i+1}\cdots
{z}^{\left(b^{(i-1)}_{s^{(i-1)}}-1\right)}_{p_0+s_0+\cdots +p_{i-1}+s_{i-1}+p_i+s^{(i-1)}}\times
\nonumber\\
\times {z}^{\left(a^{(i)}_1-1\right)}_{p_0+s_0+\cdots +p_{i-1}+s_{i-1}+p_i+s^{(i-1)}+1}\cdots
{z}^{\left(a^{(i)}_{s^{(i)}}-1\right)}_{p_0+s_0+\cdots +p_{i-1}+s_{i-1}+p_i+s_i-1}
\end{eqnarray}
for $i=(1,\dots l)$, and
\begin{eqnarray}
g_{l+1}({\bf z}) ={z}^{\left(--a_1^{(l)}\right)}_{p_1+s_1+\cdots +p_{l}+s_{l}+1}\cdots
{z}^{\left(-a^{(l)}_{p^{(l)}}\right)}_{p_1+s_1+\cdots +p_{l}+s_{l}+p^{(l)}}\times\nonumber\\
\times {z}^{\left(-a_1^{(l+1)}\right)}_{p_1+s_1+\cdots +p_{l}+s_{l}+p^{(l)}+1}\cdots
{z}^{\left(-a^{(l+1)}_{s^{(l+1)}}\right)}_{p_1+s_1+\cdots +p_{l}+s_{l}+p_{l+1}}\times\nonumber\\
\times {z}^{\left(b^{(l)}_1-1\right)}_{p_1+s_1+\cdots +p_{l}+s_{l}+p_{l+1}+1}\cdots
{z}^{\left(b^{(l)}_{s^{(l)}}-1\right)}_{p_1+s_1+\cdots +p_{l}+s_{l}+p_{l+1}+s^{(l)}}\times
\nonumber\\
\times {z}^{\left(b^{(l+1)}_1-1\right)}_{p_1+s_1+\cdots +p_{l}+s_{l}+p_{l+1}+s^{(l)}+1}\cdots
{z}^{\left(b^{(l+1)}_{s^{(l+1)}}-1\right)}_{p_1+s_1+\cdots +p_{l}+s_{l}+p_{l+1}+s_{l+1}-1}
\end{eqnarray}

\section*{Conclusion}

We get multivariable hypergeometric functions as certain tau-functions
of the KP hierarchy.
It means that we have a set of new relations on the
multivariable hypergeometric functions. 
For instance all
hypergeometric functions of the form (\ref{tauhyp1})
or of the form (\ref{gen}) satisfy bilinear Hirota equations \cite{DJKM}
of the KP hierarchy.
Hypergeometric functions may be also considered as ratios of
the tau-functions of the 
two-dimensional Toda lattice evaluated at special values of Toda lattice 
times.
To get   hypergeometric functions of a single set of arguments
(\ref{hZ1}),(\ref{vh}) one should take ${\bf t}^*$ as in (\ref{t^*}),
(\ref{chvy}).
To get   hypergeometric functions of a double set of arguments
(\ref{vh2}), (\ref{hZ}) one should keep the descrete time $M$
 constant.
One can get the fermionic representations for different special
functions and polynomials related to these hypergeometric functions.
Using integral representation one can express hypergeometric functions 
as the integral of rather simple hypergeometric function. We also get
determinant representation of (\ref{tauhyp1}), which may allow to analize
analytical properties of multivariable tau-functions in terms
of functions of only one variable. We wrote down the system of 
linear equations on tau-function (\ref{tauhyp1}), which may allow
to find applications to quantum mechanical problems. 
It is quite unexpected that we obtain $q$-deformed version of these 
hypergeometric functions as tau-functions not of a $q$-deformed KP hierarchy 
\cite{Bogd},\cite{Mir},\cite{Sem},\cite{Fren},\cite{Ming} but of the 
usual KP hierarchy. It is now an
 interesting problem to establish links between these results and
 group-theoretic approach to the 
$q$-special functions \cite{V,KV} and matrix integrals. Let us note
a certain similarity of some of our formulas and formulas from \cite{KR}.
 We expect to work out connections with matrix models of Kontsevich 
type \cite{Ko} and two-matrix models related to 2D Toda lattice 
\cite{GMMO,KMM}. 
We consider (in \cite{AO,Sc,OSc}) the multicomponent 
KP \cite{JM,KL}  as a basis for 
obtaining new examples of hypergeometric functions.

Let us note the paper \cite{O2} which turns to be
different example of the tau function (\ref{tau3}). This tau-function
is not of hypergeometric type, but is closely related to it.
This tau-function has an interesting meaning in the algebraic geometry.
In \cite{O1} the interpretation of the Schur functions as a measure
on partitions is explained, thus we have an additional interpretation
of hypergeometric functions as generating functions for certain 
probabilities described in \cite{O1}.

In the present version of the paper we add references to the
\cite{AN},\cite{Ka} where different examples of special functions
of one and of two variables were considered as solutions of
Hirota equations with variable coefficients. It will be
interesting to get the fermionic representations for these examples.

\app{Formulae involving $r(D)$}
Let $r(n)=e^{T_{n-1}-T_n}$, $h(n)=e^{T_n}$. Then the operators 
$r(D),h(D)$, where  $D=z\frac{d}{dz}$, have the  simple properties
\be
F_{r}(M,zz^*):=
\left(1+\sum_{n=1}^\infty e^{T_{M-n}-T_{M}}\frac{(zz^*)^{-n}}{n!}
\right)(zz^*)^M=
\ee
\be
=\left(1-\frac{1}{zz^*}\frac{r(D)}{D+a}\right)^{-a}\cdot (zz^*)^M=
\left(1-\frac{1}{zz^*}\frac{r(D)}{D+b}\right)^{-b}\cdot (zz^*)^M ,
\quad a,b\neq -M,
\ee
\be\label{expPDO}
\xi_r^{(\infty)}(\beta ,z,D)
:=\sum_{n=1}^\infty \beta_n \left(\frac{1}{z}r(D)\right)^n,\quad
\xi_r^{(0)}({\bf t},z,D)
:=\sum_{n=1}^\infty t_n \left(r(D)z\right)^n  .
\ee
Let us consider 
\begin{eqnarray}\label{1DGGF}
f_r(M,\beta,z):=
(zz^*)^M\left(1+\sum_{n=1}^\infty e^{T_{M-n}-T_{M}}z^{-n}p_n(\beta)\right)
=e^{\xi_r^{(\infty)}(\beta ,z,D)}\cdot (zz^*)^M.
\end{eqnarray}

Let $h(n)=e^{T_n}$. Then
\begin{eqnarray}
\frac{h(D)}{h(0)}\cdot f_r(M,\beta,z)=
(zz^*)^{M}e^{\xi(\beta,z^{-1})},\quad 
\xi({\beta},z^{-1}):=\sum_{m=1}^{\infty}\beta_m z^{-m} .
\end{eqnarray}
{\bf Remark}. Let us put $m\beta_m=a(z^*)^{-m}, m=1,2,...$ with some 
$a\neq 0$, then $ f_r(M,\beta,z)=F_{r'}(M,zz^*)$, where $r'(D)=(D+a)r(D)$,
since the elementary Schur polynomials (\ref{elSchur})
$p_n(\beta)=\frac{(a)_n}{n!}(z^*)^{-n}$. Eq.(\ref{sym=miwa})
generalizes this property for all partitions.

\app{Orthogonal polynomials and matrix integrals}
It is known that the hypergeometric functions (\ref{hZ}) appear 
in the group representation theory and are connected with 
the so-called matrix integrals \cite{V}. On the other hand
the set of examples \cite{GMMO,Ko} reveals a
connection between the matrix integrals \cite{Mehta} and the soliton theory.
To establish this connection in our case, it is useful to consider the 
related systems of the orthogonal polynomials. Let us briefly describe how 
to write down these polynomials.

Let $M_+$ be the largest integer zero of $r$. 
Then the function 
\be\label{pm}
f^+_r(zz^*)=\sum_{n=0}^{+\infty}(zz^*)^{n+M_+} e^{T_{n+M_+}-T_{M_+}}
\ee
is the eigenfunction of the operator $\frac{1}{z}r(D)$ with the eigenvalue
$z^*$. 
Since operator
$r(D)$ is invertible on the functions $\{ z^M,M>M_+\}$ we write
\be
f^+_r(zz^*)=
\left(1-\frac{1}{r(D)}zz^*\right)^{-1}\cdot (zz^*)^{M_+}.
\ee
For example if we take $r(n)=n$ we obtain 
\be
f^+_r(zz^*)=e^{zz^*}.
\ee
 We use this function as weight function for
a system of orthogonal polynomials $\{\pi_n^{\pm},n=0,1,2,\dots \}$,
related to the 
hypergeometric solution of KP:
\be\label{ortpol}
\int_\gamma \pi _n^{-}({\bf t},\beta,z)
e^{\xi({\bf t},z)}f^{+}_r(zz^*)e^{\xi({\beta},z^*)} 
\pi _m^{+}({\bf t},\beta,z^*)dzdz^*
=e^{-\phi_{M_{+}+ n}({\bf t},\beta)}\delta_{n,m} .
\ee
The corresponding two matrix integral \cite{Mehta} is the following one
\be\label{matint}
\tau(M,{\bf t},\beta)=\int e^{Tr\xi({\bf t},Z)}f^{+}_r\left(Tr(ZZ^*)\right) 
e^{Tr\xi({\beta},Z^*)}dZdZ^*.
\ee
Here $Z,Z^*$ are Hermitian $M \times M$ matrices. 

\app{The vertex operator action}
Now we present
relations between hypergeometric functions 
which follow from the soliton theory, for instance see \cite{ASM},\cite{D'}.
Let us introduce the operators which act on functions of ${\bf t}$ 
variables:
\be
\Omega_r^{(\infty)}({\bf t}^*):=-
\frac{1}{2\pi i} \lim _{\epsilon \to 0}
\oint V^*_\infty (z+\epsilon)
\xi_r^{(\infty)}({\bf t}^* ,z,D)V_\infty (z) dz ,
\ee
\be
\Omega_r^{(0)}({\bf t}):=
\frac{1}{2\pi i} \lim _{\epsilon \to 0}
\oint  V_0 (z+\epsilon)
\xi_r^{(0)}({\bf t},z,D)V_0^* (z) z^{-2} dz ,
\ee
where $V_\infty(z),V^*_\infty(z),V_0(z),V_0^*(z)$ are defined by 
(\ref{vertex}). 
For instance 
\be
\Omega_r^{(\infty)}({\bf t}^*)=\Omega_r^{(0)}({\bf t})
=\sum_{n>0} nt_nt^*_n,\quad r=1 .
\ee
The bosonization formulae \cite{JM},\cite{UT} give the relations which 
connect hypergeometric functions (\ref{tauhyp})
and hypergeometric functions (\ref{gen}):
\be
e^{-\Omega_{\tilde{r}^1}^{(\infty)}(\tilde{\g}_1)} \cdots
 e^{-\Omega_{\tilde{r}^k}^{(\infty)}(\tilde{\g}_k)}\cdots
e^{\Omega_{r^l}^{(\infty)}(\g_l)} \cdots
 e^{\Omega_{r^l}^{(\infty)}(\g_l)}
\tau_r(M,{\bf t},{\bf t}^*)=
\tau (M,{\bf t},{\bf t}^*;\tilde{\g},\g) ,
\ee
\be
 e^{\Omega_{r^l}^{(0)}(\g_l)}\cdots
 e^{\Omega_{r^l}^{(0)}(\g_l)} \cdots
 e^{-\Omega_{\tilde{r}^k}^{(0)}(\tilde{\g}_k)}\cdots
 e^{-\Omega_{\tilde{r}^1}^{(0)}(\tilde{\g}_1)}
\tau_r(M,{\bf t},{\bf t}^*)=
\tau (M,{\bf t},{\bf t}^*;\tilde{\g},\g) .
\ee
In particular we have shift argument formulae for the tau function  
(\ref{tauhyp}):
\be\label{stringvert1}
e^{\Omega_r^{(\infty)}(\gamma)}\tau_r(M,{\bf t},{\bf t}^*)=
\tau_r(M,{\bf t},{\bf t}^* +\g),\quad 
e^{\Omega_r^{(0)}(\g)}\tau_r(M,{\bf t},{\bf t}^*)=
\tau_r(M,{\bf t}+\g,{\bf t}^*) .
\ee
Also we have
\be\label{stringvert2}
e^{\sum_{-\infty}^{\infty}\g_n Z_{nn}}\tau (M,{\bf t},{\bf T},{\bf t}^*)=
e^{\sum_{-\infty}^{\infty}\g_nZ_{nn}^*}\tau (M,{\bf t},{\bf T},{\bf t}^*)=
\tau(M,{\bf t},{\bf T}+\g,{\bf t}^*),
\ee
where
\be
Z_{nn}=-
\frac{1}{4\pi^2} 
\oint \frac {z^n}{{z^*}^n}V^*_\infty (z^*)V_\infty (z) dzdz^* ,\quad
Z_{nn}^*=-
\frac{1}{4\pi^2} 
\oint \frac {z^n}{{z^*}^n}V^*_0 (z^*)V_0 (z) dzdz^* .
\ee
For instance
\be
e^{\sum_{-\infty}^{\infty}T_n Z_{nn}}
\exp \left(\sum_{n=1}^\infty nt_nt^*_n\right)=
e^{\sum_{-\infty}^{\infty}T_nZ_{nn}^*}
\exp \left(\sum_{n=1}^\infty nt_nt^*_n\right)=
\tau (M,{\bf t},{\bf T},{\bf t}^*).
\ee

\app{Gauss factorization problem, additional symmetries, string equations
and $\Psi DO$ on the circle}
Let us describe relevant string equations following Takasaki and Takebe
\cite{TT},\cite{T}. We shall also consider this topic in a more detailed 
paper.

Let us introduce infinite matrices to describe KP and TL
flows and symmetries, see \cite{UT}. Zakharov-Shabat dressing 
matrices are $K$ and $\bar{K}$. $K$ is a lower triangular matrix with unit 
main diagonal: $(K)_{ii}=1$. $\bar{K}$ is an upper triangular matrix.
The matrices $K,\bar{K}$ depend on parameters 
$M,{\bf t},{\bf T},{\bf t}^*$. 
The matrices $(\Lambda)_{ik}=\delta_{i,k-1}$,
$(\bar{\Lambda})_{ik}=\delta_{i,k+1}$.
For each value of 
${\bf t},{\bf T},{\bf t}^*$ and $M \in Z$ they solve Gauss (Riemann-Hilbert) 
factorization problem for infinite matrices: 
\begin{eqnarray}
\bar{K}=KG(M,{\bf t},{\bf T},{\bf t}^*),\quad G(M,{\bf t},{\bf T},{\bf t}^*)=
\exp \left( \xi({\bf t},\Lambda)\right)
\Lambda^{M}G({\bf 0},{\bf T},{\bf 0})\bar{\Lambda}^{M}
\exp \left(\xi({\bf t}^*,\bar{\Lambda})\right) .\quad
\end{eqnarray}
We put 
$\log (\bar{K}_{ii})=\phi_{i+M}$, and a set of fields 
$\phi_i({\bf t},{\bf t}^*), ( -\infty <i<+\infty) $ 
solves the hierarchy of higher two-dimensional TL equations.

Take 
$L=K\Lambda K^{-1}$, $\bar{L}=\bar{K}\bar{\Lambda}{\bar{K}}^{-1}$,
and
$(\Delta)_{ik}=i\delta_{i,k}$, 
$\widehat{M}=K\Delta K^{-1}+M+\sum nt_n L^n$,  
$\widehat{\bar{M}}=\bar{K}\Delta {\bar{K}}^{-1}+M+\sum nt^*_n{\bar{L}}^n$.
Then the KP additional symmetries \cite{TT},\cite{T},\cite{D'},\cite{ASM},
\cite{OW} and higher TL flows \cite{UT} are written as
\be
\partial_{\beta_n}K=
-\left(\left(r(\widehat{M})L^{-1}\right)^n\right)_-K,\quad
\partial_{\beta_n}\bar{K}=
\left(\left(r(\widehat{M})L^{-1}\right)^n\right)_+\bar{K},\quad
\ee
\be
\partial_{t^*_n}K=-\left(\bar{L}^n\right)_-K,\quad
\partial_{t^*_n}\bar{K}=\left(\bar{L}^n\right)_+\bar{K}.\
\ee
Then the string equations are
\be\label{string1}
\bar{L}L=r(\widehat{M}),
\ee
\be\label{string2}
\widehat{\bar{M}}=\widehat{M}.
\ee
The first equation is a manifestation of the fact that the
group time $\beta_1$ of
the additional symmetry of KP can be identified with the Toda lattice time
$t^*_1$. In terms of tau-function we have the equation (\ref{stringvert1}) 
in terms of vertex operator action \cite{D'},\cite{ASM}, or the 
equations (\ref{linur}) in case the tau-function is written in Miwa 
variables.\\
 The second string equation (\ref{string2}) is related to the symmetry of 
our tau-functions with respect to ${\bf t} \leftrightarrow \beta$.

When  
\be\label{rlin}
r(M)=M+a,
\ee
the equations (\ref{string1}),(\ref{string2}) describe $c=1$ string , see 
\cite{NTT},\cite{T}. In this case we easily get the relation
\be\label{barLL}
[{\bar L},L]=1 .
\ee
The string equations in the form of Takasaki
 allows us to notice the similarity to the different problem.
The dispersionless limit of (\ref{barLL})
(and also of (\ref{string2}), (\ref{string1}), where $r(M)=M^n$, and of
(\ref{rlin})) will be written as
\be
{\bar \lambda}\lambda=\mu^n,\quad n \in Z ,
\ee
\be
{\bar \mu}=\mu .
\ee
The case when ${\bar \lambda}$ and ${\bar \mu}$ are complex conjugate of
 $\lambda$ and of $\mu$ respectively, is of interest. These string equations 
(mainly the case $n=1$) were 
recently investigated to solve the so-called Laplacian 
growth problem, see \cite{MWZ}. We are grateful to A.Zabrodin for
the discussion on this problem. For the dispersionless limit of the KP and TL
hierarchies see \cite{TT}.

In case the function $r(n)$ has zeroes (described by divisor
${\bf m}=(M_1,...)$: $r(M_k)=0$), one needs to
produce the replacement:
\be
\Lambda \to \Lambda({\bf m}),\quad 
\bar{\Lambda} \to \bar{\Lambda}({\bf m}) ,
\ee
where new matrices $\Lambda({\bf m})$,$\bar{\Lambda}({\bf m})$ are defined
as
\be
(\Lambda({\bf m}))_{i,j}=\delta_{i,j-1}, j\neq M_k,\quad
(\Lambda({\bf m}))_{i,j}=0, j= M_k ,
\ee
\be
(\bar{\Lambda}({\bf m}))_{i,j}=\delta_{i,j+1}, i\neq M_k,\quad
(\bar{\Lambda}({\bf m}))_{i,j}=0, i= M_k .
\ee
This modification describes the open
TL equation (\ref{deltaToda}):
\be\label{deltaToda'}
\partial_{t_1}\partial_{\beta_1}\varphi_n=
\delta(n)e^{\varphi_{n-1}-\varphi_{n}}-
\delta(n+1)e^{\varphi_{n}-\varphi_{n+1}} . 
\ee
The set of fields $\phi_\infty,\dots ,\phi_{M_1},\phi_{M_1+1}\dots$ 
consists of the following

 parts due to the conditions
\be\label{Todaspliting}
\sum_{n=-\infty}^{M_s}\phi_n=0,\quad
\sum_{n=M_k+1}^{M_{k+1}}\phi_n=0,\quad
\sum_{n=\infty}^{M_1+1}\phi_n=0,
\ee
which result from $\tau(M_k,{\bf t},{\bf t}^*)=1$.

Each internal  part (\ref{finitechainh}) of this Toda chain
is a $M_{k+1}-M_{k}$ sites  chain. There are two semiinfinite parts
which correspond to (\ref{semiinfh}) and (\ref{semiinf'h}).

\bremark 
The matrix $r(\widehat{M})$ contains $(\widehat{M}-b_i)$
in the denominator. The matrix $(\widehat{M}-b_i)^{-1}$ 
is $K(\Delta -b_i)^{-1}K^{-1}(1+O({\bf t}))$ (compare
the consideration of the inverse operators with \cite{OW}).
\eremark

The KP tau-function (\ref{tauhyp}) can be obtained as follows.
\be
G(M,{\bf t},{\bf T},{\bf t}^*)=
G({\bf 0},{\bf T},{\bf 0})U(M,{\bf t},\beta),\quad
U(M,{\bf t},\beta)=U^{+}({\bf t})U^{-}(M,\beta ).
\ee
\be
U^{+}({\bf t})=\exp \left( \xi({\bf t},\Lambda)\right),\quad
U^{-}(M,\beta )=
\exp \left(\xi(\beta,\Lambda ^{-1}r\left(\Delta+M\right))\right),
\ee
The matrix $G({\bf 0},{\bf T},{\bf 0})$
is related to the transformation of the eq.(\ref{Toda}) 
to the eq.(\ref{rToda}).
By taking the projection \cite{UT} $U \mapsto U_{--}$ for nonpositive 
values of matrix 
indices we obtain a determinant representation of 
the tau-function (\ref{tauhyp}):
\be \label{2-c}
\tau_r(M,{\bf t},\beta)=
\frac {\det U_{--}(M,{\bf t},\beta)}
{ \det \left(U^{+}_{--}({\bf t}) \right)
\det \left(U^{-}_{--}(M,\beta ) \right)}
=\det U_{--}(M,{\bf t},\beta) ,
\ee
since both determinants in the denominator are equal to one.
Formula (\ref{2-c}) is also a Segal-Wilson formula for ${GL}(\infty)$ 
2-cocycle \cite{SW} $C_M\left( U^{+}(-{\bf t}),U^{-}(-\beta )\right)$. 
Choosing the function $r$ as in {\em Section 3.2}
we obtain hypergeometric functions listed in the {\em Introduction}.\\
\bremark
 Therefore the hypergeometric functions which were 
considered above
have the meaning of $GL(\infty)$ two-cocycle on the two 
multiparametrical group elements $U^{+}({\bf t})$ and  
$U^{-}({M,\beta})$. 
Both elements $U^{+}({\bf t})$ and  $U^{-}(M,\beta )$  can be considered as
elements of group
of pseudodifferential operators on the circle. The corresponding Lie 
algebras consist of the multiplication operators $\{z^n;n \in N_0\}$
and of the pseudodifferential operators
$\{\left( \frac 1z r(z\frac {d}{d z}+M) \right)^n;n \in N_0\}$.
Two sets of group 
times ${\bf t}$
and $\beta$ play the role
of indeterminates of the hypergeometric functions (\ref{tau}).
Formulas (\ref{tauhyp}) and (\ref{trr}) mean the expansion
of ${GL}(\infty)$ group 2-cocycle in terms of corresponding
Lie algebra 2-cocycle
\be
c_M (z,\frac 1z r(D))=r(M),\quad 
c_M (\tilde{r}(D)z,\frac 1z r(D+M))=\tilde{r}(M)r(M).
\ee
Japanese cocycle is cohomological to Khesin-Kravchenko cocycle
\cite{KhK} for the $\Psi DO$ on the circle:
\be
c_M \sim c_0 \sim \omega_M ,
\ee
which is
\be
\omega_M(A,B)=\oint res_\partial A[\log (D+M),B]dz,
\quad A,B \in \Psi DO .
\ee
For the group cocycle we have
\be\label{CM}
C_M \left(e^{-\sum z^nt_n},e^{-\sum (z^{-1}r(D))^n\beta_n}\right)
=\tau_r(M,{\bf t},\beta ),
\ee
where we imply that the order of $\Psi$DO $r(D)$ is 1 or less.
About properties of $e^{-\sum (z^{-1}r(D))^n\beta_n}$ see 
(\ref{expPDO}),(\ref{1DGGF}).
\eremark
\bremark
It is interesting to note that in case of hypergeometric functions
${}_pF_s$ (\ref{hZ1}) the order of $r$ is $p-s$ (see {\em Example 3}), 
and the condition
$p-s \leq 1$ is the condition of the convergence of this hypergeometric 
series, see \cite{KV}. Namely the radius of convergence is 
finite in case $p-s=1$, it is infinite when $p-s<1$ and it is 
zero for $p-s>1$ (this is true for the case when no one of $a_k$
in (\ref{hZ1}) is nonnegative integer).
\eremark
\bremark
The set of functions $\{ w(n,z), n=M,M+1,M+2,\dots \}$, where
\be\label{SatoGr}
w(n,z)= 
\exp \left( -\sum_{m=0}^\infty t^*_m\left(\frac 1z r(D)\right)^m\right)
\cdot z^n ,
\ee
may be identified with Sato Grassmannian (\ref{SatoG}) related to the 
cocycle (\ref{CM}). 
The dual Grassmannian (\ref{dSatoG}) is the set of one forms
$\{ w^*(n,z),n=M,M+1,M+2,\dots \}$,
\be\label{dSatoGr}
 w^*(n,z)=\exp \left( \sum_{m=0}^\infty t^*_m\left(\frac 1z r(-D)\right)^m
\right)\cdot z^{-n} dz .
\ee
\eremark

\app{Equations with respect to ${\bf T}$ variables}

Let us show that variables ${\bf T}$ play the role of time variables.
We shall write down some equation involving the differentiation
with respect to ${\bf T}$. The relevant Liouville equations
is constructed in terms of
\be
e^{v_{k}}=\frac{
\l M+1|e^{H({\bf t})}\psi_ke^{H_0({\bf T})}
e^{H^*({\bf t}^*)}|M\r
\l M-1|e^{H({\bf t})}\psi^*_k
e^{H_0({\bf T})}
e^{H^*({\bf t}^*)}|M\r}
{\l M|e^{H({\bf t})}e^{H_0({\bf T})}e^{H^*({\bf t}^*)}|M\r^2} .
\ee
The equations are
\be
-\frac{\partial^2v_k}{\partial t_1 \partial T_k}=e^{v_k},\quad k \in Z.
\ee

To get three-wave equations take
\be
\beta_{nm}=\frac
{\l M|e^{H({\bf t})}\psi_m\psi^*_ne^{H_0({\bf T})}
e^{H^*({\bf t}^*)}|M\r}
{\l M|e^{H({\bf t})}e^{H_0({\bf T})}e^{H^*({\bf t}^*)}|M\r} ,
\ee
\be
\beta_{1'n}=\frac
{\l M+1|e^{H({\bf t})}\psi_ne^{H_0({\bf T})}
e^{H^*({\bf t}^*)}|M\r}
{\l M|e^{H({\bf t})}e^{H_0({\bf T})}e^{H^*({\bf t}^*)}|M\r},
\quad
 \beta_{n1'}=\frac
{\l M-1|e^{H({\bf t})}\psi^*_ne^{H_0({\bf T})}
e^{H^*({\bf t}^*)}|M\r}
{\l M|e^{H({\bf t})}e^{H_0({\bf T})}e^{H^*({\bf t}^*)}|M\r} .
\ee
Then we obtain 
\be
-\partial_{T_n}\beta_{1'm}=\beta_{1'n}\beta_{nm},\quad
-\partial_{T_n}\beta_{m1'}=\beta_{mn}\beta_{n1'},\quad
\partial_{t_1}\beta_{mn}=\beta_{m1'}\beta_{1'n},\quad m\neq n,
\quad n,m \in Z.
\ee

One obtains these equations with the help of the  Lax type
representation \cite{Or}: 
\be
[\partial_{T_m}+
\beta_{1'm}\partial_{t_1}^{-1}\beta_{m1'},
\partial_{T_n}+\beta_{1'n}\partial_{t_1}^{-1}\beta_{n1'}]=0.
\ee
It is possible to write down the discrete versions of these equations,
 and (the discrete and the continues) equations involving only  ${\bf T}$ 
variables.

\app{Orthogonal $q$-polynomials}
Now we present the fermionic representation of polynomials listed in 
the {\em Introduction}. These polynomials are obtained by the specification
of (\ref{qssN=1}).\\
{\bf $q$-Askey-Wilson polynomials} are defined as
\begin{eqnarray}
p_n(x;a,b,c,d|q)=a^{-n}(ab;q)_n(ac;q)_n(ad;q)_n \nonumber\\
\times {}_4\varphi_3\left.\left(q^{-n}, q^{n-1}abcd, ae^{i\eta}, 
ae^{-i\eta}\atop ab,\quad ac,\quad ad\right| q,q\right), \qquad x=\cos \eta .
\end{eqnarray}
Let operator ${}_4r^{(q)}_3(D)$ be
\be\label{rq-askey}
{}_4r_3^{(q)}(D)= 
\frac{(1-q^{-n+D})(1-abcdq^{n-1+D})(1-ae^{i\eta}q^D)(1-ae^{-i\eta}q^D)}
{(1-abq^{D})(1-acq^D)(1-adq^D)} .
\ee
For this operator we have:
\be
\frac{{}^4\tau^3(M,{\bf t},{\bf T},{\bf t}^*)}
{{}^4\tau^3(M,{\bf 0},{\bf T},{\bf 0})}=
\l M|e^{H({\bf t})}e^{-A({\bf t}^*)}|M\r ,
\ee
where
\be
A_k=\frac{1}{2\pi i}
\oint \psi^*(z) \left(\frac{1}{z}{}_4r_3^{(q)}(D)\right)^k \psi(z),
\quad k=1,2,\dots .
\ee
If we  take the variables ${\bf t}^*$ as in (\ref{chvy}) and 
\be\label{qxvar}
{\bf t}=(q,\frac{q^2}{2},\frac{q^3}{3},\dots) ,
\ee 
\begin{eqnarray}\label{q-askey}
\frac{{}_4\tau_3^{(q)}(M,{\bf t},{\bf T},{\bf t}^*)}
{{}_4\tau_3^{(q)}(M,{\bf 0},{\bf T},{\bf 0})}={}_4\varphi_3\left.
\left(q^{M-n}, q^{M+n-1}abcd, aq^Me^{i\eta}, aq^Me^{-i\eta}\atop q^Mab,
\quad q^Mac,\quad q^Mad\right| q,q\right)
=\nonumber\\ \sum_{m=0}^{+\infty}\frac{(q^{M-n};q)_{m}
(q^{M+n-1}abcd;q)_m(aq^Me^{i\eta};q)_m(aqe^{-i\eta};q)_m}{(abq^{M};q)_{m}
(acq^M;q)_m (adq^{M};q)_{m}}\frac{q^{m}}{(q;q)_m}.
\end{eqnarray}
For $M=0$ we obtain $q$-Askey-Wilson polynomials:
\begin{eqnarray}
p_{n}(x;a,b,c,d|q)=aq^{-n}(ab;q)_{n}(ac;q)_{n}(ad;q)_{n}
\frac{{}^4{\tau}^3(0,{\bf t},{\bf T},{\bf t}^*)}
{{}^4{\tau}^3(0,{\bf 0},{\bf T},{\bf 0})} .
\end{eqnarray}
If parameters in (\ref{rq-askey}) are
\be
a=q^{\frac{2\alpha+1}{4}},\quad b=-q^{\frac{2\nu+1}{4}},
\quad c=q^{\frac{2\alpha+3}{4}},\quad d=-q^{\frac{2\nu+3}{4}} ,
\ee
we get a fermionic representation for continuous $q$-{\bf Jacobi polynomials}
\be
P_{n}^{(\alpha,\nu)}(x|q)=\frac{(q^{\alpha+1};q)_{n}}{(q;q)_{n})}
\frac{{}^4\tau^3(0,{\bf t},{\bf T},{\bf t}^*)}
{{}^4\tau^3(0,{\bf 0},{\bf T},{\bf 0})} .
\ee
For  $c=-a, b=-d=q^{\frac{1}{2}}a$ we have $q$-{\bf Gegenbauer polynomials}
\be
C_{n}(\cos \eta;\mu|q)=\frac{(\mu^2;q)_{n}}{\mu^{\frac{n}{2}}(q;q)_{n}}
\frac{{}^4\tau^3(0,{\bf t},{\bf T},{\bf t}^* )}
{{}^4\tau^3(0,{\bf 0},{\bf T},{\bf 0} )},\quad \mu=a^2 .
\ee  
{\bf Clebsch-Gordan coefficients} 
$C_q({\bf l},{\bf j})$ see \cite{V}.\\
Let
\be
{}_3r_2^{(q)}(D)= \frac{(1-q^{j-l_1+D})(1-q^{l_1+j+1+D})(1-q^{-l+m+D})}
{(1-q^{l_2-l+j+1+D})(1-q^{-l-l_2+j+D})} .
\ee
For variables from (\ref{chvy}) and (\ref{qxvar}) we have
\be
\frac{{}_3\tau^{(q)}_2(M,{\bf t},{\bf T},{\bf t}^*)}
{{}_3\tau^{(q)}_2(M,{\bf 0},{\bf T},{\bf 0})}={}_3\Phi_2\left.
\left(j-l_1+M, l_1+j+1+M, -l+m+M\atop l_2-l+1+M, -l-l_2+j+M\right|q,q\right) .
\ee
Thus we have the  fermionic representation 
\begin{eqnarray}
C_q({\bf l},{\bf j})=\frac{(-1)^{l_1-j}q^B\Delta({\bf l})[l+l_2-j]!([{\bf l},
{\bf j}][2l+1])^{\frac{1}{2}}}{[l_1-l_2+l]![l+l_2-l_1]![l_2-l+j]![l_1-j]!
[l_2+k]![l-m]!} \frac{{}_3\tau^{(q)}_2(0,{\bf t},{\bf T},{\bf t}^*)}
{{}_3\tau^{(q)}_2(0,{\bf 0},{\bf T},{\bf 0})}.\qquad
\end{eqnarray}
{\bf $q$-Hahn polynomials}\\
Let us take the  operator:
\be
{}_3r_2^{(q)}(D)= \frac{(1-q^{-n+D})(1-abq^{n+1+D})(1-q^{-x+D})}
{(1-aq^{D+1})(1-q^{D-N})},\quad n\le N .
\ee
The corresponding tau function whose variables are defined in (\ref{chvy}) 
and (\ref{qxvar}):
\be
\frac{{}_3\tau_2^{(q)}(M,{\bf t},{\bf T},{\bf t}^*)}
{{}_3\tau_2^{(q)}(M,{\bf 0},{\bf T},{\bf 0})}=
{}_3\varphi_2\left.\left(q^{M-n},
abq^{M+n+1}, q^{M-x}\atop aq^{M+1},\quad q^{M-N}\right|q,q\right).
\ee
Therefore  the fermionic representation of $q$-Hahn polynomials is
\be
Q_{n}(q^{-x};a,b;-N|q)=\frac{{}_3\tau_2^{(q)}(0,{\bf t},{\bf T},{\bf t}^*)}
{{}_3\tau_2^{(q)}(0,{\bf 0},{\bf T},{\bf 0})}.
\ee
{\bf $q$-Racah polynomials}
Now we take as $r(D)$ 
\be\label{rq-racah}
{}_4r_3^{(q)}(D)= 
\frac{(1-q^{-n+D})(1-abq^{n+1+D})(1-q^{-x+D})(1-cdq^{x+1+D})}
{(1-aq^{D+1})(1-bdq^{D+1})(1-cq^{D+1})} .
\ee
The tau function is:
\be
\frac{{}_4\tau^{(q)}_3(M,{\bf t},{\bf T},{\bf t}^*)}
{{}_4\tau^{(q)}_3(M,{\bf 0},{\bf T},{\bf 0})}=
{}_4\varphi_3\left.\left(q^{M-n}, abq^{M+n+1}, q^{M-x}, cdq^{M+x+1}
\atop aq^{M+1},\quad bdq^{M+1},\quad cq^{M+1}\right|q,q\right) ,
\ee
where ${\bf t}^*$ and ${\bf t}$ as in (\ref{chvy}) and (\ref{qxvar}).
 Thus we have an expression
\be
R_{n}(\mu(x); a, b, c, d|q)=
\frac{{}_4\tau_3^{(q)}(0,{\bf t},{\bf T},{\bf t}^*)}
{{}_4\tau_3^{(q)}(0,{\bf 0},{\bf T},{\bf 0})},
\quad \mu(x)=q^{-x}+cdq^{x+1}.
\ee
 {\bf Little $q$-Jacobi polynomials}\\
Setting operator $r(D)$:
\be
{}_2r_1^{(q)}(D)=\frac{(1-q^{-n+D})(1-abq^{n+1+D})}{(1-aq^{D+1})},
\ee
we get tau function:
\be
\frac{{}_2\tau_1^{(q)}(M,{\bf t},{\bf T},{\bf t}^*)}
{{}_2\tau_1^{(q)}(M,{\bf 0},{\bf T},{\bf 0})}=
{}_2\varphi_1\left.\left(q^{M-n},abq^{M+n+1}\atop aq^{M+1}
\right| q,qx\right),
\ee
where $t_m=\frac{(qx)^m}{m}$ and ${\bf t}^*$ as in (\ref{chvy}).
\be
p_n(x;a,b|q)=\frac{{}_2\tau_1^{(q)}(0,{\bf t},{\bf T},{\bf t}^*)}
{{}_2\tau_1^{(q)}(0,{\bf 0},{\bf T},{\bf 0})}.
\ee

\app{}
Since we have 
$\exp\left( T_n:\psi_n^*\psi_n:\right)=
1+\left(e^{T_n}-1\right):\psi_n^*\psi_n:$ 
the tau-function (\ref{tau3}) is a linear function of each
$e^{T_m}-1$. Take in the series (\ref{tau3}) a coefficient 
before the monomial $\prod_i (e^{T_{m_i}}-1)$, where $\{m_i\}$ form a set of 
indices $X$. Denote this coefficient as  
$\rho (X,{\bf t},{\bf t}^*)\exp\left(-\sum_{k=1}^\infty kt_kt_k^*\right)$.  
The function $\rho (X,{\bf t},{\bf t}^*)$ has the following meaning in 
the probability  theory \cite{O2}. The Schur measure on 
partitions is defined via the formula:
\be\label{Schurmeasure}
\mu ({\bf n},{\bf t},{\bf t}^*)=s_{\bf n}({\bf t})s_{\bf n}({\bf t}^*)
\exp\left(-\sum_{k=1}^\infty kt_kt_k^*\right),
\ee
where ${\bf n}$ is a partition, and ${\bf t},{\bf t}^*$ 
are parameters. Then the function $\rho (X,{\bf t},{\bf t}^*)$ 
describes the probability that the set 
$\{n_i -i\}$ contains a given set $X \subset Z$. The Schur measure has
a natural group-theoretical interpretation, see \cite{O2}. In this context it
may be of interest to consider the following
measure on partitions, which generalizes (\ref{Schurmeasure}):
\be
\mu_{\tilde{r}r} ({\bf n},{\bf t},{\bf t}^*)=\frac 
{{\tilde{r}}_{ \bf n }(M)r_{ \bf n }(M)s_{\bf n }({\bf t})
s_{\bf n }({\bf t}^* ) }
{\l M |e^{\tilde{A}({\bf t})}e^{A({\bf t}^*)} |M \r },
\ee
see (\ref{trr}). Let us note that in the case of hypergeometric 
tau-functions, function
$r_{ \bf n }$ is a rational function of the Schur polynomials, see 
(\ref{qtaushur}),(\ref{taushur}) below.

\section*{Acknowledgements}
One of the authors (A.O.) thanks Vl.Dragovich and T.Shiota 
for the helpful numerous discussions. D.S. would like to thank S.Senchenko for 
helpful discussions. 
We thank S.Milne for sending \cite{Milne}, L.A.Dickey for \cite{DK},
 J.Harnad and A.Its for the interesting 
remarks after the first author's talk at the CRM workshop ``Isomonodromy 
deformations'', May 2000.
We thank also A.Zabrodin for the discussion of the paper \cite{MWZ}.
We thank A. Nakamura for sending \cite{AN}.

\end{document}